\documentclass{aa}
\usepackage{psfig}
\usepackage{lscape}
\usepackage{amsmath}
\usepackage{amssymb}
\usepackage{latexsym}
\usepackage[figuresleft]{rotating}

\newcommand{\MC}{\multicolumn}
\newcommand{\kms}{km~s$^{-1}$}
\newcommand{\sunn}{$_{\odot}$}
\newcounter{qub}
\begin{document}

\title{HS 0837+4717 -- a  metal-deficient blue compact galaxy
with large nitrogen excess
}

\author{S.Pustilnik\inst{1,7} \and
A.Kniazev\inst{1,2,7} \and
A.Pramskij\inst{1,7} \and
Y.Izotov\inst{3} \and
C.Foltz\inst{4} \and
N.Brosch\inst{5} \and
J.-M.Martin\inst{6} \and
A.Ugryumov\inst{1,7}
}
\offprints{S. Pustilnik, \email{sap@sao.ru}}

\institute{
Special Astrophysical Observatory RAS, Nizhnij Arkhyz,
Karachai-Circassia,  369167 Russia
\and Max Planck Institut f\"{u}r Astronomie, K\"{o}nigstuhl 17,
D-69117,
Heidelberg, Germany
\and Main Astronomical Observatory, 27 Zabolotnoho str., Kyiv, 03680,
Ukraine
\and National Science Foundation, 4201 Wilson Blvd., Arlington,
Virginia, 22230, USA
\and the Wise Observatory and the Raymond and Beverly Sackler Faculty
of
Exact Sciences, Tel-Aviv University, Tel Aviv 69978, Israel
\and Observatoire de Paris, place J. Janssen, F-92195 Meudon cedex, France
\and Isaac Newton Institute of Chile, SAO Branch
}

 \date{Received \hskip 1cm 7 November, 2003; accepted \hskip 1cm 18 February 2004}

 \abstract{
We present the results of high S/N long-slit spectroscopy with the
Multiple
Mirror and the SAO 6-m telescopes, optical imaging with the Wise 1-m
telescope
and \ion{H}{i} observations with the Nan\c {c}ay Radio Telescope
 of the very metal-deficient
(12+$\log$(O/H)=7.64) luminous ($M_{\rm B}$=--18\fm1) blue compact
galaxy (BCG)  HS~0837+4717. The
blue bump near $\lambda$4650,
characteristic of WR stars
is detected in the central supergiant \ion{H}{ii} region, as well as
the barely seen red bump at $\lambda$5808.
The derived number of WR stars in the region of the current starburst is
$\sim$1000. Evidence for fast motions in this region is also seen as broad,
low-contrast components in the H$\alpha$, H$\beta$ and strong
[\ion{O}{iii}]
lines $\lambda\lambda$4959,5007. While the extinction of the narrow
emission
lines from the supergiant \ion{H}{ii} region is low, the very large
Balmer decrement of the broad
components suggests that the part of current starburst is highly
obscured by dust.
Abundance ratios X/O for X=Ne, Ar, S, Fe and Cl in the supergiant \ion{H}{ii}
region are in good agreement with the mean values of other very
metal-deficient BCGs.
Nitrogen, however, is
overabundant by a factor of $\sim$6.
This implies an unusually efficient N enrichment in HS 0837+4717, and
probably, a non-typical evolution scenario. The H$\alpha$-line
Position--Velocity
(P--V) diagrams for directions approximately along the major and minor
axes reveal disturbed
motions of the ionized gas, mainly in peripheral regions.
The SW part of the major axis P--V diagram looks like a rotation curve,
with
the velocity amplitude $V_{\rm rot} \sim$50--70~\kms\ at $r$=4.3 kpc. Its
NE part displays, however, strong deviations, indicating either
counter-rotation, or a strong outflow/supershell.
If it is considered as indicating a
shell-like feature
its velocity amplitude of $\sim$70
\kms\ (relative to the extrapolated rotation curve), and the apparent
extent
of $\sim$4\arcsec\ (3.3 kpc) imply a dynamical age of $\sim$14 Myr and the
full energetic equivalent of $\sim$2.3$\times$10$^4$ SNe. The latter
indicates continuing starbursts during at least this time interval.
The long-slit spectra reveal a complex morphology for this galaxy. It
consists of two compact regions at a distance of $\sim$2 kpc. Their
continuum flux
differs by a factor of four. The brightest one is related to the
current starburst with the age of $\sim$3.7 Myr.
The slightly redder
fainter component could be an older starburst ($\sim$25
Myr).  The  Wise 1-m telescope $UBVR$ integrated photometry reveals
a high optical luminosity for this BCG, and the unusual $(B-V)$
and $(V-R)$ colours. The morphology of HS 0837+4717 is highly disturbed,
with two small tails emerging to NNW and SSE.
Such a disturbed overall morphology, a "double-nucleus" structure,
significantly disturbed velocities of ionized gas, together with the
very high
power of the starburst suggests a possible explanation of the object as
a recent merger.
We compare the properties of this BCG and of
similar objects
known in the
literature, and conclude that their high nitrogen excess is most
probably related to the short phase of a powerful starburst
when many WR stars contribute to the enrichment of ISM.
   }

\authorrunning{S.Pustilnik et al.}

\titlerunning{Study of the very metal-deficient BCG HS~0837+4717}

\maketitle

\section{Introduction}

Most low-mass gas-rich galaxies have low metallicities, with the
typical range of 1$Z$\sunn/15 to $Z$\sunn/3  (e.g., review of Kunth \&
\"Ostlin \cite{Kunth2000}).
Only for a very small fraction of galaxies in the Local Universe is the
metallicity extremely low, in the
range of $Z$\sunn/50 to $Z$\sunn/20. This range is more
characteristic of high-redshift damped Ly$\alpha$ 
absorption systems which could be young galaxies.
The study of the extremely metal-deficient (XMD) galaxies
allows one to test many
theoretical ideas on massive star formation and evolution, and models of
galaxy evolution in a very low metallicity environment.

In particular,
the abundance patterns of $\alpha$-elements and iron, that is, the
ratios X/O (with X=Ne, S, Ar, Fe), allow one to check models of stellar
nucleosynthesis. For metal-poor galaxies they were found to be remarkably
constant over a wide range of O/H (e.g., Izotov \& Thuan \cite{IT99},
hereafter IT99).
This implies that all these elements are primary
and are produced in the same massive stars. The ratio N/O was also found to
be fairly constant for XMD BCGs
(IT99), implying mainly primary production of nitrogen
in very low metallicity environment.

Wolf-Rayet (WR) stars -- a specific very short phase of massive star
evolution, characterized by strong mass outflow, and enriched mainly in
N and C, are often seen in high S/N spectra of starburst galaxies (e.g.,
Kunth \& Joubert \cite{KJ85}; Izotov et al. \cite{Izotov96};
Guseva et al. \cite{Guseva2000}).
These stars are detected through the broad emission features of the so
called ``blue bump'' near $\lambda$4650~\AA, and more rarely in the red,
near $\lambda$5808~\AA.
Models of stellar evolution predict that the number of WR stars in a young
star cluster is a sensitive function of metallicity, IMF and  starburst
age (e.g., Schaerer \& Vacca \cite{SV98}, hereafter SV98).

WR star outflows could produce a significant N
overabundance at the locations of young starbursts. However, the
analysis of a large sample of \ion{H}{ii} galaxies by
Kobulnicky \& Skillman (\cite{KS96})
showed that galaxies, with and without strong WR features in their
integrated spectra, show identical N/O ratios.
This is probably explained by nitrogen outflow in the hot phase.
Therefore, only significantly after its ejection could the nitrogen
be cooled and mixed into the optically observed ionized gas.

Significant nitrogen overabundance is detected in
a few starburst galaxies. The most prominent one is Mkn~996
(Thuan et al. \cite{Thuan96}), with 12+$\log$(O/H)=8.0. It is
a non-typical BCG due to the large
number of WR stars and to a powerful outflow from the compact central
SF burst. Its N/O in the region with diameter of 0.6 kpc is 4 to 25 times
higher than the typical value for low-metallicity BCGs
of $\sim$1/40.
The morphology of Mkn 996 implies that likely it is a
remnant of a recent merger. The authors
argue that the observed large N/O ratio is directly related to the
powerful outflows of numerous WR stars.

Another well known case is a factor of three nitrogen overabundance,
found in two compact regions ($d \sim$20 pc) near the central
starburst in the nearby ($D\sim$4~Mpc) dwarf WR galaxy NGC~5253 (e.g.,
Kobulnicky et al. \cite{Kobul97}, and the summary of previous results
therein).

Only three galaxies:
UM~420, Mkn ~1089, and UM~448, all luminous and with detected WR lines
(Guseva et al. \cite{Guseva2000}), from the sample of 50
BCGs in IT99, show nitrogen
overabundance of a factor of $\approx$3 relative to the mean N/O value of
1/30 for their range of O/H). Finally, the galaxy Haro 11 from the sample
of luminous BCGs (Bergvall \& \"Ostlin \cite{Bergvall02}), also with strong
WR bump, shows an N/O ratio that is 6 times higher than for the main
BCG group.

In this paper we present the results of a comprehensive study of
HS~0837+4717,  found in the Hamburg/SAO survey
(Pustilnik et al. \cite{Pustilnik99}).
The very low metallicity of HS 0837+4717 ($Z\approx$$Z$\sunn/20) was
first
derived from the observations with the SAO RAS 6-m telescope (BTA) in
1996 (Kniazev et al. \cite{Kniazev00a}).
Classified as an XMD BCG, it is unusually luminous for
its metallicity. This galaxy shows WR bumps and broad
components of strong emission lines of hydrogen and oxygen.
This is the next most nitrogen overabundant galaxy known after Mkn
996. But, in contrast to the latter, its optical spectrum does not look
atypical for BCGs, and its very large nitrogen overabundance is seen on
an otherwise more or less typical BCG background.

In Sect. \ref{Obs} we describe the observations and their reduction.
In Sect. \ref{Results} all the results from the reduction and
preliminary
analysis are presented, including the heavy element abundances, the WR
and broad components of strong lines, the ionized gas kinematics and the
morphology and photometry.
In Sect. \ref{discussion} we discuss
the obtained results and derived
parameters of this galaxy, compare its properties
with those of other luminous and
XMD  BCGs, and draw conclusions.
The adopted distance to the galaxy is 170.5 Mpc; respective scale is
827 pc per arcsecond.

\section{Observations and data reduction}
\label{Obs}

\subsection{Spectrophotometry and kinematic data}
\label{Spectral}


\begin{table*}
\begin{center}
\caption{Journal of spectral observations of HS~0837+4717}
\label{Tab1}
\begin{tabular}{lrrrccccc} \\ \hline \hline
\MC{1}{c}{ Telescope}     &
\MC{1}{c}{ Date }       &
\MC{1}{c}{ Exposure }   &
\MC{1}{c}{ Wavelength } &
\MC{1}{c}{ Dispersion } &
\MC{1}{c}{ Seeing }     &
\MC{1}{c}{ Airmass }     &
\MC{1}{c}{ PA }          \\

\MC{1}{c}{ }       &
\MC{1}{c}{ }       &
\MC{1}{c}{ time [s] }    &
\MC{1}{c}{ Range [\AA] } &
\MC{1}{c}{ [\AA/pixel] } &
\MC{1}{c}{ [arcsec] }    &
\MC{1}{c}{          }    &
\MC{1}{c}{ [degree] }     \\

\MC{1}{c}{ (1) } &
\MC{1}{c}{ (2) } &
\MC{1}{c}{ (3) } &
\MC{1}{c}{ (4) } &
\MC{1}{c}{ (5) } &
\MC{1}{c}{ (6) } &
\MC{1}{c}{ (7) } &
\MC{1}{c}{ (8) } & \\
\hline
\\[-0.3cm]
MMT & 01.05.1997  & 2$\times$1200 & $ 3600-8300$  & 1.9 &  1.6 & 1.13 & ~~~16 \\
BTA & 19.02.2002  & 2$\times$1800 & $ 3700-6100$  & 2.4 &  2.2 & 1.27 & ~~~44 \\
BTA & 19.02.2002  & 2$\times$1800 & $ 6000-7200$  & 1.2 &  1.3 & 1.15 & 44,130 \\
\hline \hline \\[-0.2cm]
\end{tabular}
\end{center}
\end{table*}

   \begin{figure}
   \centering
   \includegraphics[angle=-0,width=8cm]{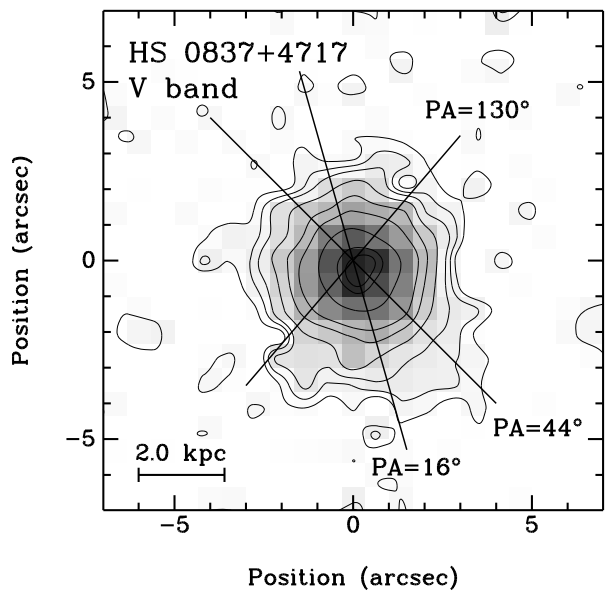}
    \caption{Morphology of HS 0837+4717 as seen in the Wise $V$-band
       image. North is up, East is to the left.
       The positions of MMT (PA=16\degr) and BTA (PA=44\degr\
        and 130\degr) long slits are shown.
          The isophotes sample the image at the brightness levels of
          11, 20, 30, 50, 75, 100, 150, 200 and 220 units.
	  The minimal level corresponds to $\sim$1.5$\sigma$ of the noise.
	  See Sect. \ref{imaging}.
              }
         \label{FigDirect}
   \end{figure}

Spectrophotometric observations of HS~0837+4717 with high S/N ratio
were obtained with the MMT on May 1, 1997.
The observations were performed with the Blue Channel of the MMT
Spectrograph using a highly-optimized Loral 3072$\times$1024 CCD
detector.
A 1\farcs5$\times$180\arcsec\ slit was used along with a 500 grooves
mm$^{-1}$ grating used in the first order and with the L-38 second
order blocking filter.
This yields a spatial scale along the slit of 0\farcs3 pixel$^{-1}$, a
scale
perpendicular to the slit of 1.9~\AA\ pixel$^{-1}$, a spectral range
3600--8300~\AA\ and a spectral resolution of $\sim$ 7~\AA\ (FWHM).  For
these observations the
CCD rows were binned by a factor of 2, yielding a final spatial
sampling of
0\farcs6 pixel$^{-1}$. The observations cover the full spectral range in
a single frame which contains all the lines of interest. Furthermore,
the
spectra have sufficient spectral resolution to separate [\ion{O}{iii}]
$\lambda$4363 from the nearby H$\gamma$ and to distinguish between the
narrow nebular and the broad WR emission lines. The total exposure time
was 40 minutes, split into 2 equal subexposures, to allow for more
effective cosmic ray removal.
Both exposures were taken at an airmass of $\sim$1.13.
The slit was
oriented in the direction with the position angle PA = 16\degr, along
the major axis of the brightest region of the galaxy (Fig. \ref{FigDirect}).
The spectrophotometric standard star HZ~44 (Massey et al.
\cite{Massey88}) was observed for flux calibration.
Reference spectra of He--Ne--Ar were obtained before and after each
observation to provide wavelength calibration.

Observations with the BTA were conducted on February 19,
2002 with the Long-slit spectrograph (Afanasiev et al.
\cite{Afanasiev95})
at the prime focus of telescope. A CCD detector Photometrics PM1024
with 24$\times$24~$\mu$m pixel was used
for data acquisition. For abundance study a grating with 651 grooves
mm$^{-1}$ was used in the first order, giving 2.38~\AA\ pixel$^{-1}$
(FWHM=7~\AA)
and allowing the simultaneous coverage of the range 3700--6100~\AA. Two
30-min exposure spectra were averaged to yield the final spectrum.
PA=44\degr\ was selected to be
close to the apparent major axis of the object's fainter isophotes.
To complement the study of emission lines in the blue, and to study the
ionized gas kinematics, we obtained two spectra in the red with a
grating with 1302 grooves
mm$^{-1}$, giving 1.2~\AA\ pixel$^{-1}$, and FWHM=3.8~\AA. These
spectra covered the range
6000--7200~\AA, and were obtained also with the slit direction along
PA=44\degr, and in the nearly perpendicular direction of
PA=130\degr.
The long slit of 2\arcsec$\times$120\arcsec\ was used for
both observations.


   \begin{figure*}
   \vspace*{5cm}
   \centering
   \includegraphics[angle=-90,width=12cm,bb=540 40 280 800]{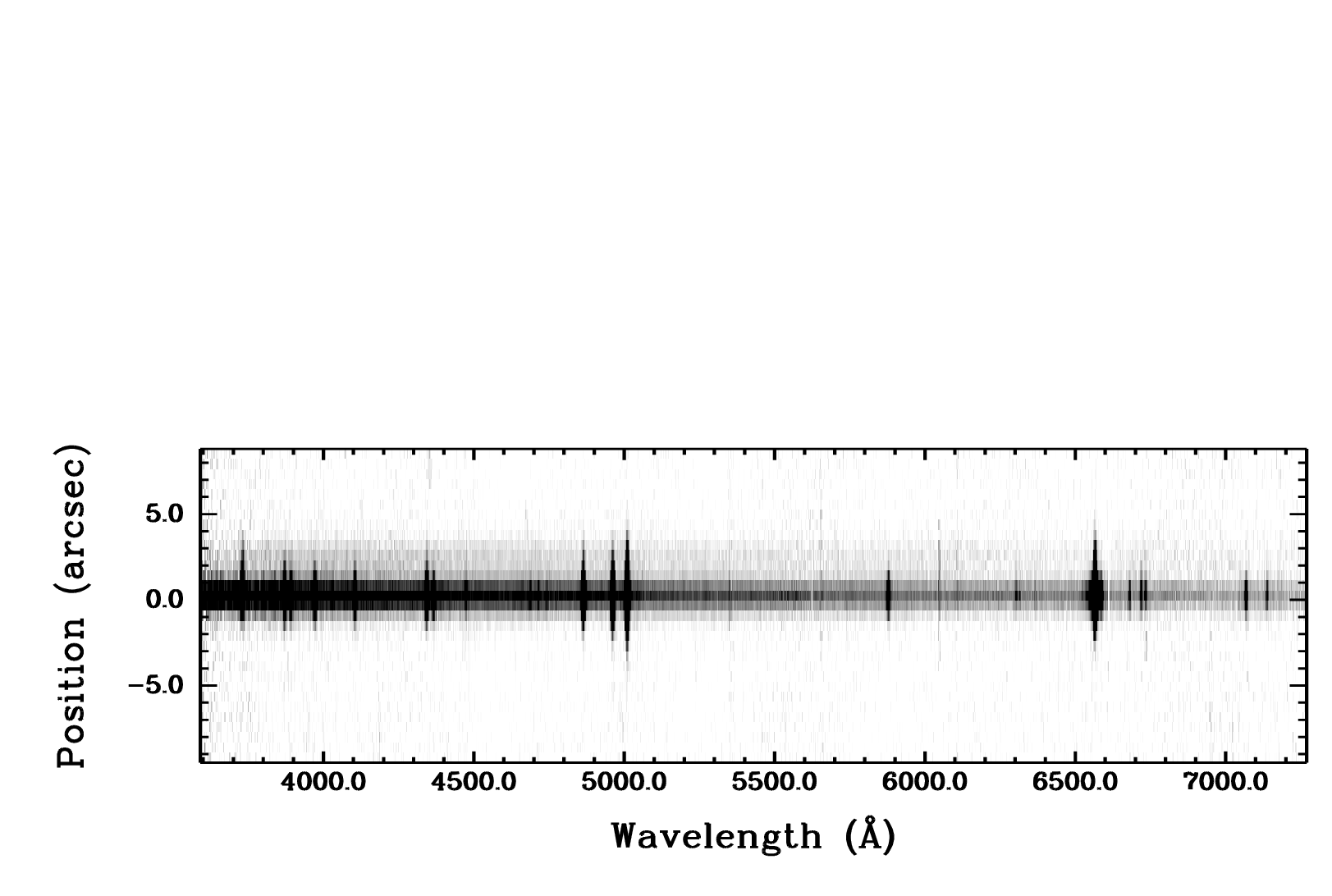}
   \includegraphics[angle=-90,width=12cm]{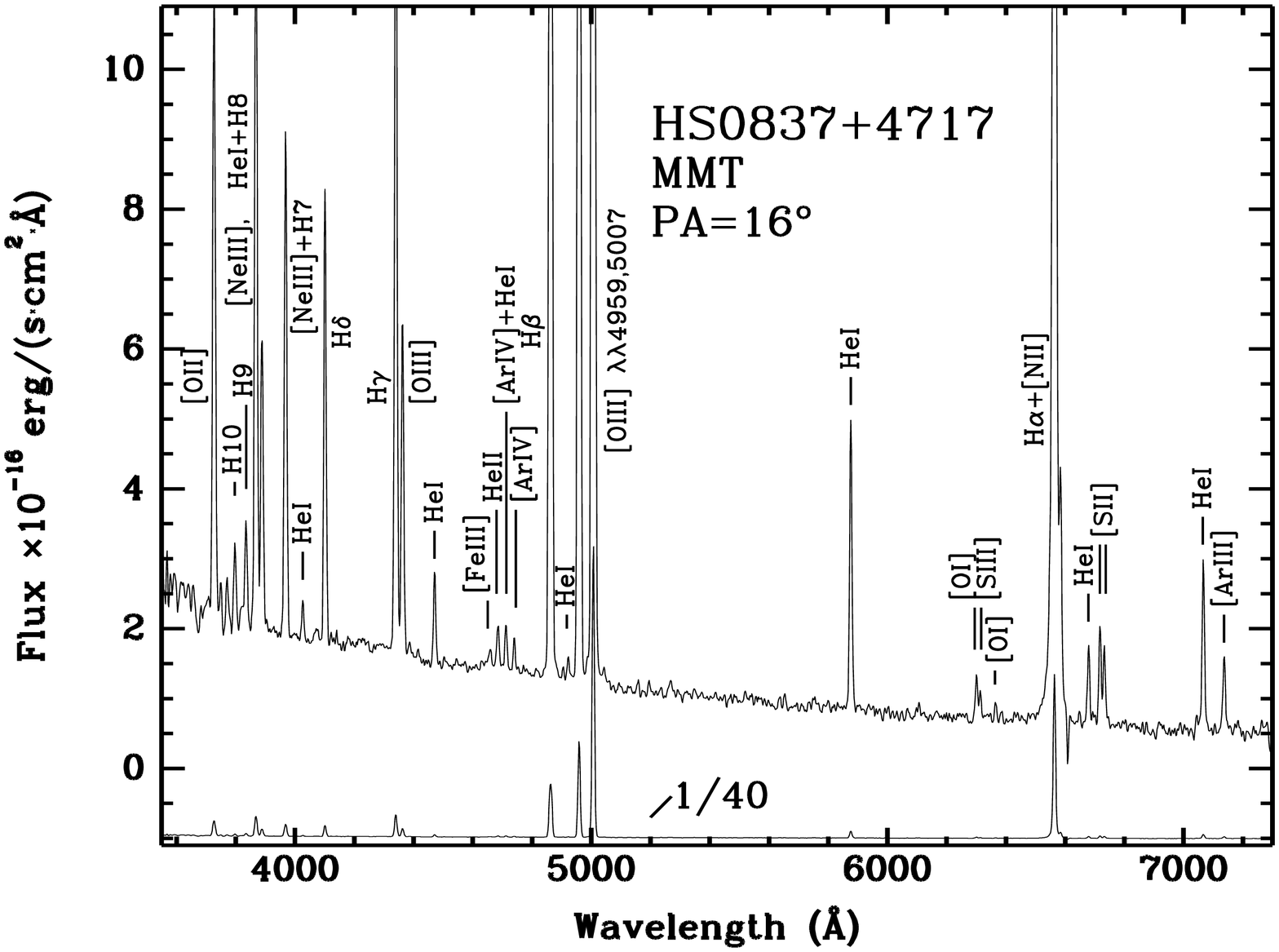}
   \includegraphics[angle=0,width=7.cm]{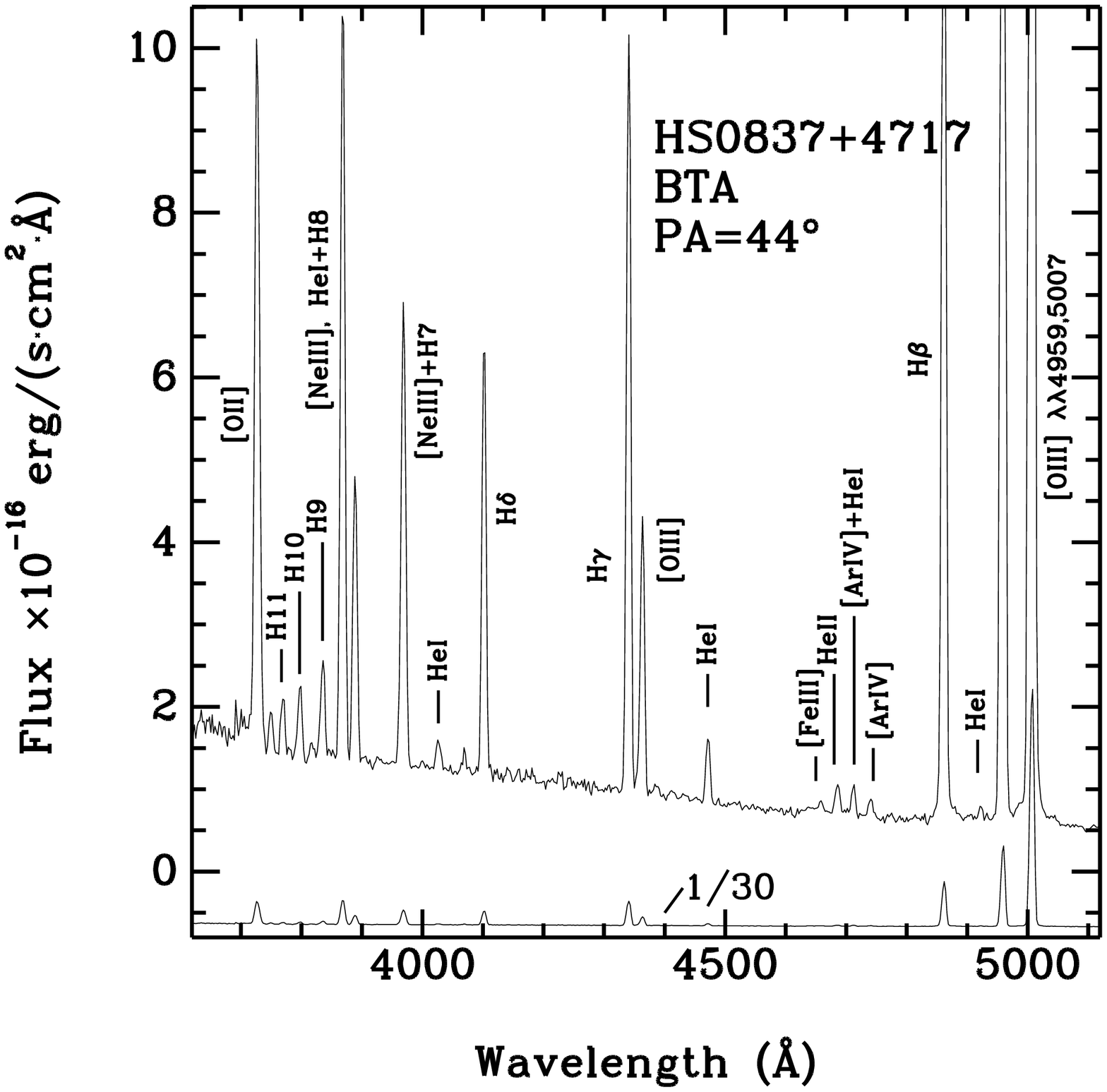}
   \includegraphics[angle=0,width=7.cm]{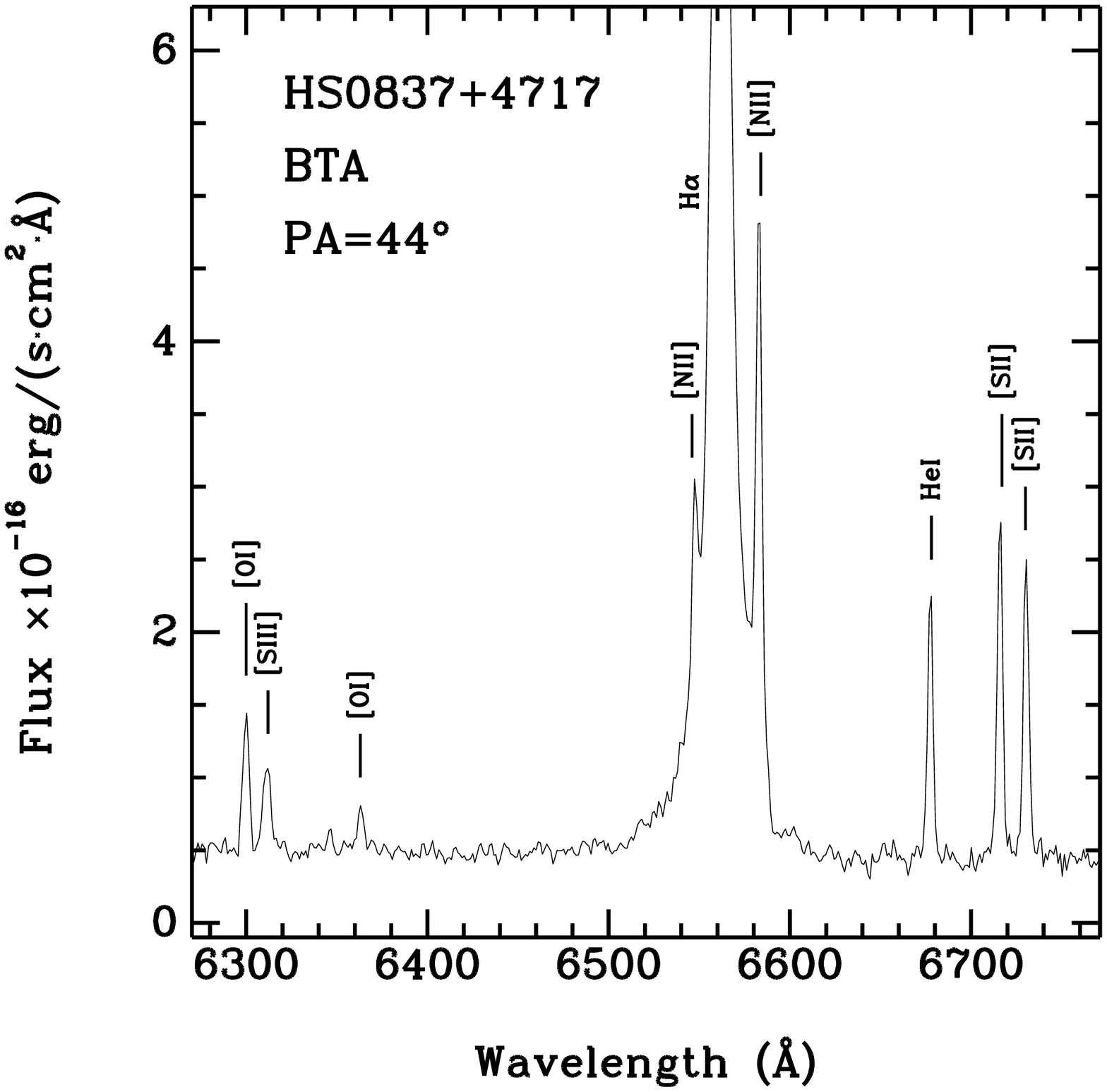}
      \caption{{\it Upper panel:} The MMT 2D spectrum of HS 0837+4717
    with PA=16\degr, plotted with the contrast level adjusted so as to
    show the distribution of the continuum intensity along the slit. SSW
    is up. While the intensity of strong lines decreases
    symmetrically relative to the continuum brightness peak, the
    continuum
        intensity at the lower level is skewed to the SSW direction,
        indicating a slightly non-central position of the starburst.
        {\it Middle panel:} De-redshifted MMT 1D spectrum of the bright
	\ion{H}{ii} region in HS 0837+4717, extracted with the aperture of
	    1\farcs5$\times$3\farcs0.
        {\it Bottom panel:} De-redshifted 1D blue and red BTA spectra.
          Note the good consistency between the MMT and BTA spectra.
              }
         \label{Fig1DSpectra}
   \end{figure*}

The reduction of the spectral observations was carried out at SAO using
the
IRAF\footnote{IRAF: the Image Reduction and Analysis Facility is
distributed by the National Optical Astronomy Observatory, which is
operated by the Association of Universities for Research in Astronomy,
In. (AURA) under cooperative agreement with the National Science
Foundation (NSF). }   and the
MIDAS\footnote{MIDAS is an acronym for the European Southern
Observatory package -- Munich Image Data Analysis System. }
software packages.
It included bias subtraction, cosmic ray removal, and flat-field
correction using exposures of a quartz incandescent lamp.
After wavelength mapping, night sky background subtraction, and
correcting
for atmospheric extinction, each frame was calibrated to absolute
fluxes. One-dimensional spectra were extracted by summing up, without
weighting,
different numbers of rows along the slit depending on exact region of
interest.
The MMT spectrum of the brightest region of HS~0837+4717 in the range
3600~\AA\ $< \lambda < $ 7300~\AA\ for the aperture
1\farcs5$\times$3\farcs0
and the BTA spectra in the blue and red, extracted with the aperture
2\arcsec$\times$2\farcs8, are shown in Fig.~\ref{Fig1DSpectra}.

\begin{table*}[hbtp]
\centering{
\caption{Line intensities in the supergiant \ion{H}{ii} region}
\label{t:Intens}
\begin{tabular}{lccccc} \hline  \hline
\rule{0pt}{10pt}
& \MC{2}{c}{MMT (1\farcs5$\times$3\farcs0)}   && \MC{2}{c}{BTA(2\arcsec$\times$2\farcs8)}  \\ \cline{2-3} \cline{5-6}
\rule{0pt}{10pt}
$\lambda_{0}$(\AA) Ion                    &
$F$($\lambda$)/$F$(H$\beta$)&$I$($\lambda$)/$I$(H$\beta$) &&
$F$($\lambda$)/$F$(H$\beta$)&$I$($\lambda$)/$I$(H$\beta$) \\ \hline
3727\ [O\ {\sc ii}]\                      & 0.380$\pm$0.019 & 0.402$\pm$0.022 && 0.379$\pm$0.021 & 0.407$\pm$0.024   \\
3835\ H9\                                 & 0.052$\pm$0.005 & 0.082$\pm$0.009 && 0.050$\pm$0.005 & 0.078$\pm$0.009   \\
3868\ [Ne\ {\sc iii}]\                    & 0.418$\pm$0.021 & 0.436$\pm$0.024 && 0.398$\pm$0.023 & 0.423$\pm$0.026   \\
3889\ He\ {\sc i}\ +\ H8\                 & 0.147$\pm$0.011 & 0.181$\pm$0.015 && 0.145$\pm$0.012 & 0.178$\pm$0.016   \\
3967\ [Ne\ {\sc iii}]\ +\ H7\             & 0.269$\pm$0.013 & 0.304$\pm$0.016 && 0.257$\pm$0.014 & 0.295$\pm$0.017   \\
4026\ He\ {\sc i}\                        & 0.018$\pm$0.003 & 0.019$\pm$0.003 && 0.018$\pm$0.003 & 0.019$\pm$0.003   \\
4069\ [S\ {\sc ii}]\                      & 0.010$\pm$0.003 & 0.010$\pm$0.003 && 0.012$\pm$0.003 & 0.013$\pm$0.003   \\
4101\ H$\delta$\                          & 0.230$\pm$0.011 & 0.260$\pm$0.013 && 0.221$\pm$0.012 & 0.255$\pm$0.015   \\
4340\ H$\gamma$\                          & 0.466$\pm$0.021 & 0.490$\pm$0.023 && 0.449$\pm$0.023 & 0.480$\pm$0.026   \\
4363\ [O\ {\sc iii}]\                     & 0.171$\pm$0.008 & 0.172$\pm$0.009 && 0.163$\pm$0.010 & 0.166$\pm$0.010   \\
4471\ He\ {\sc i}\                        & 0.047$\pm$0.004 & 0.047$\pm$0.004 && 0.037$\pm$0.003 & 0.037$\pm$0.003   \\
4658\ [Fe\ {\sc iii}]\                    & 0.004$\pm$0.002 & 0.004$\pm$0.002 && 0.003$\pm$0.001 & 0.003$\pm$0.001   \\
4686\ He\ {\sc ii}\                       & 0.019$\pm$0.002 & 0.019$\pm$0.002 && 0.019$\pm$0.002 & 0.019$\pm$0.002   \\
4701\ [Fe\ {\sc iii}]\                    & 0.001$\pm$0.001 & 0.001$\pm$0.001 && 0.004$\pm$0.002 & 0.004$\pm$0.002   \\
4713\ [Ar\ {\sc iv]}\ +\ He\ {\sc i}\     & 0.023$\pm$0.002 & 0.023$\pm$0.002 && 0.022$\pm$0.002 & 0.022$\pm$0.002   \\
4740\ [Ar\ {\sc iv]}\                     & 0.015$\pm$0.002 & 0.015$\pm$0.002 && 0.015$\pm$0.002 & 0.015$\pm$0.002   \\
4861\ H$\beta$\                           & 1.000$\pm$0.045 & 1.000$\pm$0.047 && 1.000$\pm$0.058 & 1.000$\pm$0.059   \\
4907\ [Fe\ {\sc iv}]\                     & 0.006$\pm$0.002 & 0.006$\pm$0.002 && 0.007$\pm$0.003 & 0.007$\pm$0.003   \\
4922\ He\ {\sc i}\                        & 0.012$\pm$0.002 & 0.011$\pm$0.002 && 0.014$\pm$0.003 & 0.013$\pm$0.003   \\
4959\ [O\ {\sc iii}]\                     & 2.029$\pm$0.089 & 1.950$\pm$0.089 && 1.966$\pm$0.099 & 1.923$\pm$0.099   \\
4989\ [Fe\ {\sc iii}]\                    & 0.005$\pm$0.002 & 0.005$\pm$0.002 && 0.007$\pm$0.003 & 0.006$\pm$0.003   \\
5007\ [O\ {\sc iii}]\                     & 6.089$\pm$0.268 & 5.834$\pm$0.266 && 5.900$\pm$0.300 & 5.751$\pm$0.298   \\
5041\ [Fe\ {\sc iv}]\                     & 0.007$\pm$0.002 & 0.006$\pm$0.002 && 0.010$\pm$0.003 & 0.010$\pm$0.003   \\
5047\ He\ {\sc i}\                        & 0.004$\pm$0.002 & 0.004$\pm$0.002 && 0.003$\pm$0.003 & 0.003$\pm$0.003   \\
5158\ [Fe\ {\sc ii}]\                     & 0.006$\pm$0.001 & 0.005$\pm$0.001 && 0.004$\pm$0.002 & 0.004$\pm$0.002   \\
5199\ [N\ {\sc i}]\                       & 0.006$\pm$0.002 & 0.005$\pm$0.002 && 0.005$\pm$0.002 & 0.005$\pm$0.002   \\
5271\ [Fe\ {\sc iii}]\                    & 0.009$\pm$0.003 & 0.008$\pm$0.003 && 0.012$\pm$0.003 & 0.011$\pm$0.003   \\
5518\ [Cl\ {\sc iii}]\                    & 0.004$\pm$0.001 & 0.004$\pm$0.001 && ...             & ...               \\
5538\ [Cl\ {\sc iii}]\                    & 0.003$\pm$0.002 & 0.003$\pm$0.002 && ...             & ...               \\
5755\ [N\ {\sc ii}]\                      & 0.009$\pm$0.003 & 0.009$\pm$0.002 && ...             & ...               \\
5876\ He\ {\sc i}\                        & 0.140$\pm$0.009 & 0.127$\pm$0.009 && 0.140$\pm$0.008 & 0.130$\pm$0.008   \\
6300\ [O\ {\sc i}]\                       & 0.024$\pm$0.004 & 0.022$\pm$0.004 && 0.021$\pm$0.004 & 0.019$\pm$0.004   \\
6312\ [S\ {\sc iii}]\                     & 0.011$\pm$0.004 & 0.010$\pm$0.004 && 0.016$\pm$0.004 & 0.014$\pm$0.003   \\
6364\ [O\ {\sc i}]\                       & 0.008$\pm$0.002 & 0.007$\pm$0.002 && 0.006$\pm$0.002 & 0.005$\pm$0.002   \\
6548\ [N\ {\sc ii}]\                      & 0.027$\pm$0.002 & 0.024$\pm$0.002 && 0.027$\pm$0.003 & 0.024$\pm$0.003   \\
6563\ H$\alpha$\                          & 3.126$\pm$0.139 & 2.756$\pm$0.137 && 3.075$\pm$0.156 & 2.758$\pm$0.154   \\
6584\ [N\ {\sc ii}]\                      & 0.081$\pm$0.012 & 0.071$\pm$0.011 && 0.080$\pm$0.011 & 0.072$\pm$0.010   \\
6678\ He\ {\sc i}\                        & 0.035$\pm$0.003 & 0.031$\pm$0.003 && 0.034$\pm$0.003 & 0.030$\pm$0.003   \\
6717\ [S\ {\sc ii}]\                      & 0.043$\pm$0.004 & 0.038$\pm$0.003 && 0.046$\pm$0.004 & 0.041$\pm$0.004   \\
6731\ [S\ {\sc ii}]\                      & 0.037$\pm$0.004 & 0.032$\pm$0.003 && 0.040$\pm$0.004 & 0.035$\pm$0.004   \\
7065\ He\ {\sc i}\                        & 0.081$\pm$0.005 & 0.070$\pm$0.005 && ...             & ...               \\
7136\ [Ar\ {\sc iii}]\                    & 0.040$\pm$0.004 & 0.034$\pm$0.003 && ...             & ...               \\
C(H$\beta$)\ dex          & \MC {2}{c}{0.12$\pm$0.06} && \MC {2}{c}{0.12$\pm$0.07}  \\
EW(abs)\ \AA\             & \MC {2}{c}{3.75$\pm$0.41} && \MC {2}{c}{3.55$\pm$0.42}  \\
$F$(H$\beta$)$^a$\        & \MC {2}{c}{258$\pm$3}     && \MC {2}{c}{313$\pm$9}      \\
EW(H$\beta$)\ \AA\        & \MC {2}{c}{ 230$\pm$ 6}   && \MC {2}{c}{ 222$\pm$ 9}    \\
\hline  \hline
\MC{5}{l}{$^a$ in units of 10$^{-16}$ ergs\ s$^{-1}$cm$^{-2}$.}\\
\end{tabular}
 }
\end{table*}

\subsection{Photometry data}
\label{photo}

Photometric observation of HS 0837+4717 were performed with the Wise
observatory 1-m telescope on nights 27 and 28.05.1999. in $U,B,V$ and
$R_{c}$
bands. The images were exposed through the standard filters onto a
Tektronics 1024$\times$1024 pixels CCD and
11\farcm9$\times$11\farcm9 field-of-view (at 0.696 arcsec
pixel$^{-1}$).
During the first night, the observations were performed at airmasses of
1.69 till 2.3, beginning from the $U$-band, and ending with the
$R_{c}$-band. The
exposure time was 30 min for $U$, and 15 min for the others filters.
For the
second night, the observations were collected at the airmasses from
1.54 to
2.20, with the same exposure times. Photometric standards from the
fields of
Landolt (\cite{Landolt92}) were observed at an airmass range close to
that of the studied galaxy. The seeing during both nights was
$\sim$(2--2.5)\arcsec.

The reduction of photometry was performed in IRAF, following the
standard pipe-line, including the de-biasing, cosmic ray removal and
flat-fielding. Aperture
photometry was performed in circular apertures, and the asymptotic
values of
the growth curves were accepted as the final integrated magnitudes.
The nights were not perfect; some cirrus clouds presumably were
present. Therefore, the maximal
differences, obtained for the magnitudes in these 2 nights reach 0\fm3
(in the $V$-band).
We combined the results for both nights in the attempt to
obtain
more reliable magnitudes of this galaxy. We also used as an independent
control, the $BVR$ magnitudes and colours derived from the convolution
of the MMT spectrum with the $B,V$ and $R$ passbands.

\subsection{\ion{H}{i} data}
\label{21cm}

\ion{H}{i} line observations were carried out in February 1998 with the
300m Nan\c {c}ay\footnote{The Nan\c {c}ay
Radioastronomy Station is part of the Paris Observatory and is operated
by the Minist\`ere de l'Education Nationale and Institut des Sciences
de l'Univers of the Centre National de la Recherche Scientifique.}
radio telescope (NRT). The NRT  has a half-power beam width of
3\farcm7 (EW) $\times$ 22\arcmin\ (NS) at the declination
Dec. = 0$^\circ$.

Since HS~0837+4717 had a known optical redshift,
we split the 1024-channel autocorrelator in two halves and used
a dual-polarization receiver  to increase the S/N ratio.
Each correlator segment covered a 6.4 MHz bandwidth, corresponding to
a 1350 km s$^{-1}$ velocity coverage, and was centered at the frequency
corresponding to the optical redshift.
The channel spacing was 2.6 km s$^{-1}$ before smoothing and the
effective resolution after averaging pairs of adjacent channels and Hanning
smoothing was 10.6 km s$^{-1}$. The system temperature of the receiver
was $\approx$ 40~K in the horizontal and vertical  linear polarizations.
For the target galaxy declination the telescope gain was 0.88~K/Jy.
The observations were made in the standard total
power (position switching) mode with 1-minute on-source
and 1-minute off-source integrations.
The data were reduced using the NRT standard packages,
written by the observatory's staff. The Horizontal and Vertical linear
polarization spectra
were calibrated and processed independently, and were averaged
together.

   \begin{figure}
   \centering

\includegraphics[angle=-90,width=9.cm,clip=]{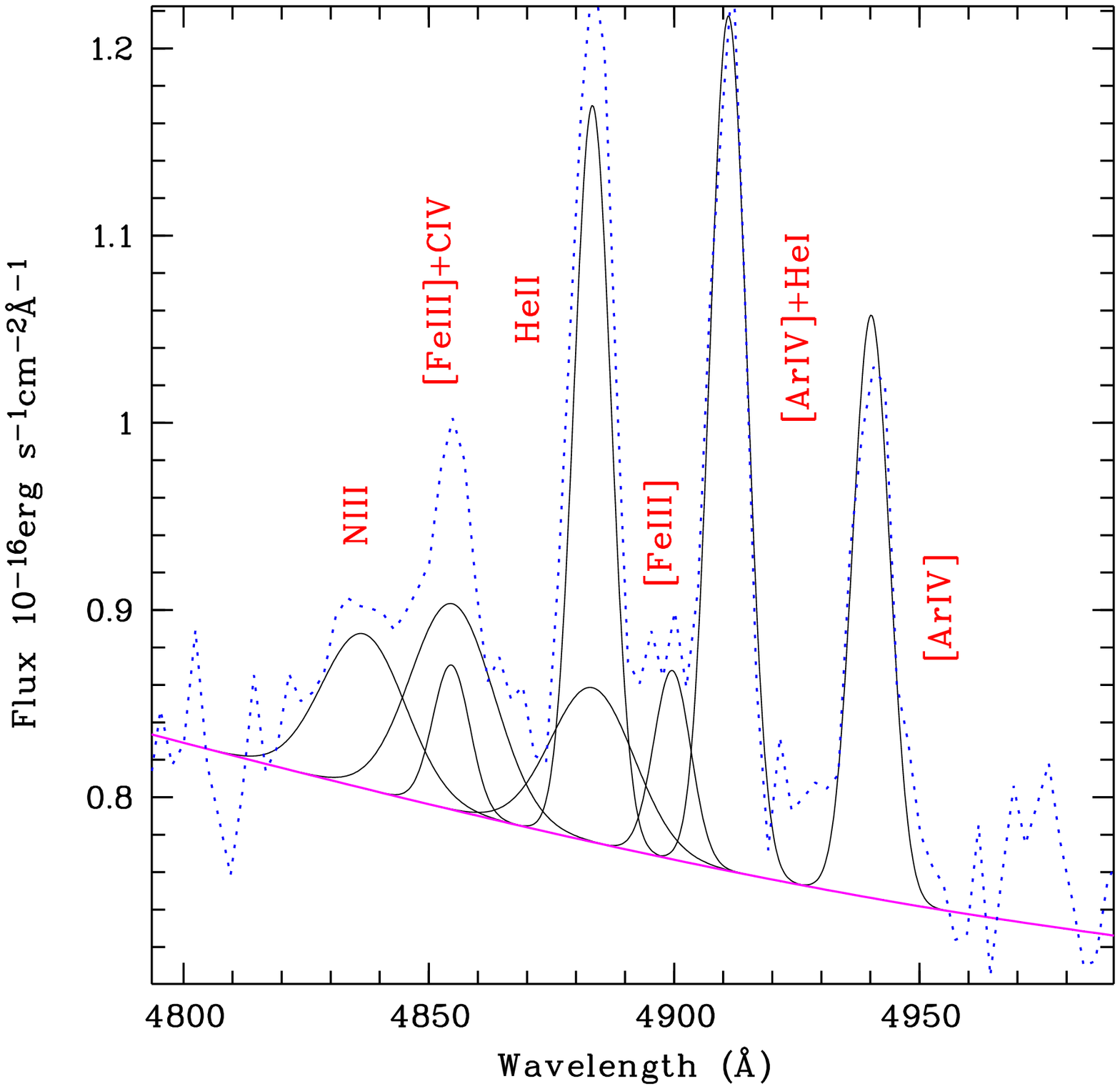}
\includegraphics[angle=-90,width=9.cm,clip=]{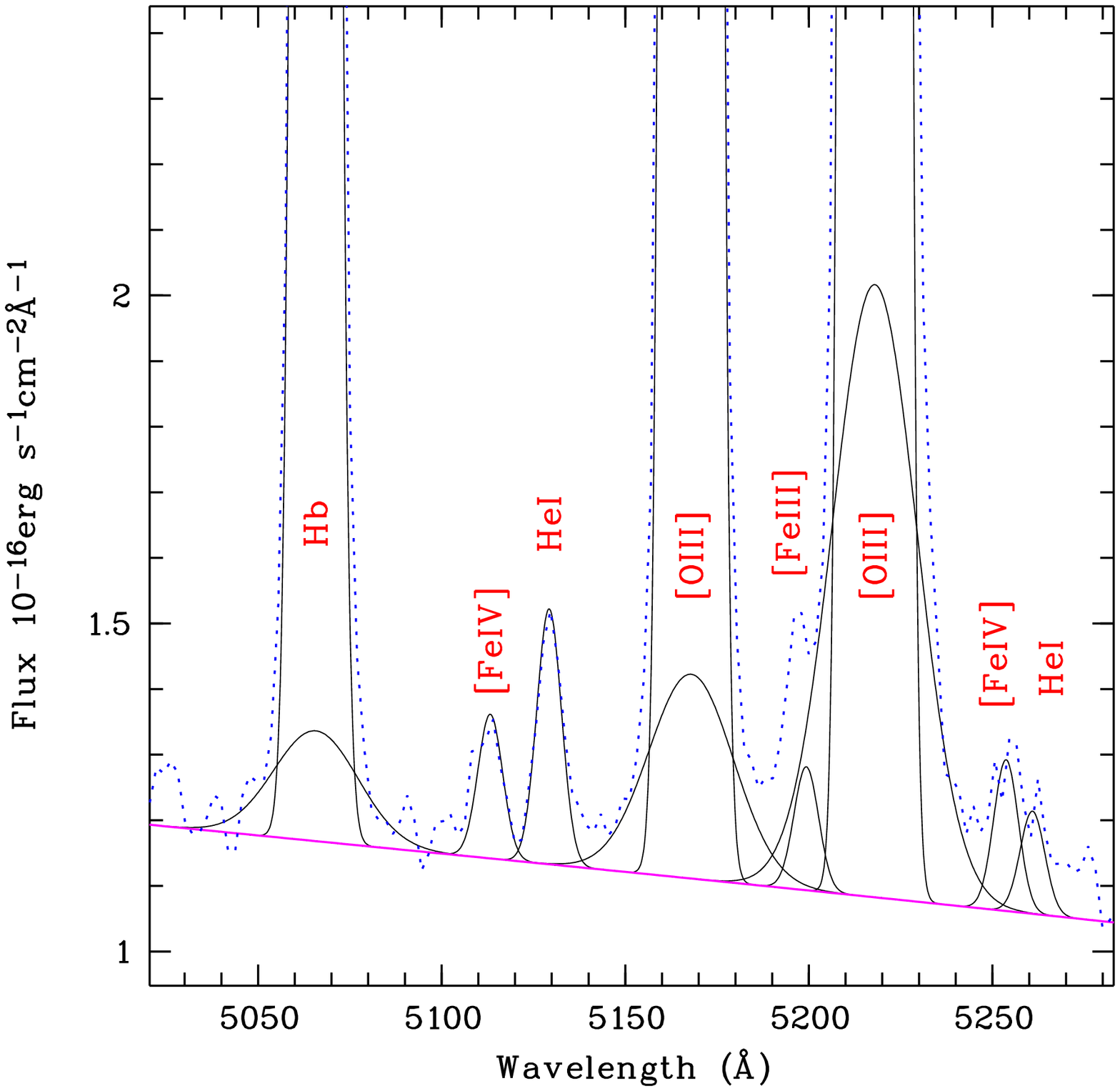}
\includegraphics[angle=-90,width=9.cm,clip=]{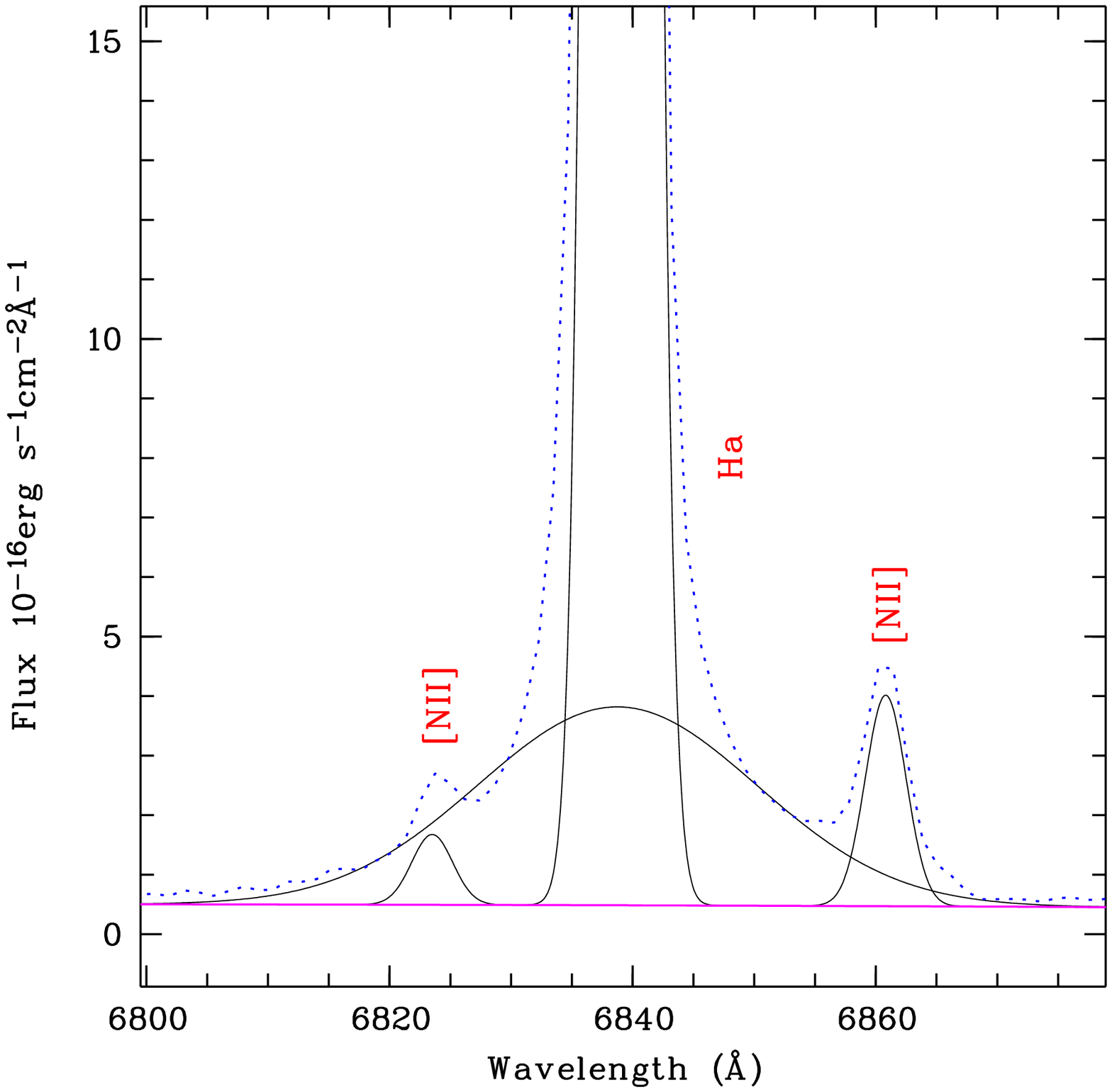}
\caption{
Gaussian multicomponent fitting (solid lines) of the BTA spectra of
HS 0837+4717 (dotted lines).
{\it top panel}: region of the blue bump;
{\it middle panel}: region of H$\beta$ and [\ion{O}{iii}]
$\lambda\lambda$4959,5007;
{\it bottom panel}: region of H$\alpha$, [\ion{N}{ii}]
$\lambda\lambda$6548,6583.
}
         \label{FigBroad}
   \end{figure}

   \begin{figure}
   \centering
   \includegraphics[angle=-0,width=6.1cm]{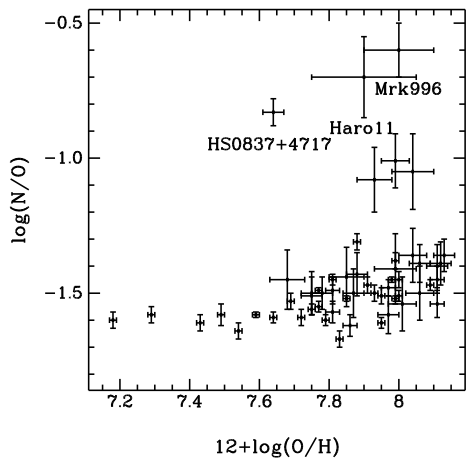}
      \caption{Distribution of $\log$(N/O) ratio versus 12+$\log$(O/H)
      for the BCG sample from Izotov \& Thuan (\cite{IT99}) with the value
      for HS~0837+4717 shown. Data for Haro 11 and Mkn 996 are from
      Bergvall \& \"Ostlin (\cite{Bergvall02}) and Thuan et al.
      (\cite{Thuan96}), respectively. The three galaxies with $\log$(N/O)
      in the range --1.1 to --1.0 are UM 420, UM 448 and Mkn 1089.
              }
         \label{fig_pattern}
   \end{figure}

\section{Results}
\label{Results}

\subsection{Chemical abundances}

To derive element  abundances of species O, Ne, N, S, Ar and Cl we use
the standard method from Aller (\cite{Aller84}), and follow the procedure
described in detail by Pagel et al. (\cite{Pagel92}) and
Izotov et al. (\cite{Izotov94}, hereafter ITL94).
The electron temperature $T_{\rm e}$ is known to be different in high-
and low-ionization \ion{H}{ii} regions (Stasi\'nska \cite{Stas90}). In the
following we have chosen to determine $T_{\rm e}$(\ion{O}{iii}) from the
[\ion{O}{iii}]$\lambda$4363/($\lambda$4959+$\lambda$5007)
ratio using the five-level atom model of \ion{O}{iii}, and
N$_{\rm e}$(\ion{S}{ii}) from the
[\ion{S}{ii}]$\lambda$6717/$\lambda$6731 ratio. We then adopt
$T_{\rm e}$(\ion{O}{iii}) for derivation of O$^{2+}$, Ne$^{2+}$ and
Ar$^{3+}$.
To derive the electron temperature for O$^{+}$ ion, we  used the
relation between  $T_{\rm e}$(\ion{O}{ii}) and $T_{\rm e}$(\ion{O}{iii})
fitted by ITL94 for the photoionized \ion{H}{ii} models of Stasi\'nska
(\cite{Stas90}). $T_{e}$(\ion{O}{ii}) has been used to
derive the O$^{+}$, N$^{+}$ and Fe$^{+}$ ionic abundances. For
Ar$^{2+}$
and S$^{2+}$ we have used an electronic temperature intermediate
between
$T_{\rm e}$(\ion{O}{ii}) and $T_{\rm e}$(\ion{O}{iii}) following the
prescription of Garnett (\cite{Gar92}).

Total element abundances have been derived after correction for unobserved
stages of ionization as described by ITL94. For the case applicable to
HS~0837+4717, when the strong nebular \ion{He}{ii} $\lambda$4686
emission
line is present, this implies that a non-negligible amount of O$^{3+}$
ion should present. Its abundance is derived from the relation:
\begin{equation}
{\rm O}^{3+} = {\rm He}^{2+}/{\rm He}^{+} ({\rm O}^{+} + {\rm O}^{2+}).
\end{equation}

Then, the total oxygen abundance is equal to
\begin{equation}
{\rm O} = {\rm O}^{+} + {\rm O}^{2+} + {\rm O}^{3+}.
\end{equation}

In Table~\ref{t:Intens} we present parameters of the
\ion{H}{ii} region on the emission extracted with
the aperture 1\farcs5$\times$3\farcs0 on the MMT spectrum,
and with the aperture 2\arcsec$\times$2\farcs8 -- on the BTA spectra.
They include both the observed line intensities and intensities
corrected
for interstellar extinction, relative to that of H$\beta$ along with
the equivalent width EW(H$\beta$), observed flux of the H$\beta$
emission
line, and extinction coefficient $C$(H$\beta$). To correct for extinction
we used the Whitford (\cite{Whitford58})  reddening law for our Galaxy.

The derived values of $T_{\rm e}$(\ion{O}{iii}), $T_{\rm e}$(\ion{O}{ii}),
$T_{\rm e}$(\ion{S}{iii}) and N$_{\rm e}$(\ion{S}{ii}) for this
supergiant \ion{H}{ii} region are shown in Table~\ref{t:Chem}.
The ionic and total abundances in the Table are derived from the two
MMT spectra according to the methods described in ITL94 and in Izotov
et al. (\cite{ITL97}, hereafter ITL97).  Since they showed a mutual
consistency, we averaged their derived parameters.

In the derivation of the O, N and Fe  abundances, the broad low
intensity underlying components of the strongest emission lines
H$\alpha$, H$\beta$
and [\ion{O}{iii}]$\lambda\lambda$4959,5007, and WR broad lines were
taken into account (see Sect. \ref{WR_lines}).
For the abundance determination only
the fluxes of the narrow (instrumental width) components of these
strong lines
were used. In the derivation of the Cl abundance, we used
the respective formula from Aller (\cite{Aller84}) with the ICF adopted
from Stasi\'nska (\cite{Stas90}). The abundances found here are well
consistent with the previous results from BTA those in Kniazev et al.
(\cite{Kniazev00a}) and those based on the SDSS spectra (Kniazev
et al. \cite{Kniazev_SDSS}).

As evident from Table \ref{t:Chem}, the
object belongs to the group of the most metal-poor
\ion{H}{ii}-galaxies, which includes only 1--2 per cent of the known BCGs.
Notice the very high abundance of nitrogen. The N/O ratio
is $\sim$6 times higher than the typical value for all
known XMD BCGs. As shown in IT99,
BCGs with 12+$\log$(O/H)$\lesssim$7.60 have $\log$(N/O)
\mbox{$\approx$--1.60} with a very small scatter (Fig. \ref{fig_pattern}).
The relative abundances of the other elements in HS 0837+4717 are close
to the mean values for the XMD BCGs (IT99).

\subsection{WR features and broad components of strong emission lines}
\label{WR_lines}

The Wolf-Rayet features of the blue bump (near $\sim$4650~\AA) and broad
low-contrast components of the strong emission lines H$\beta$, H$\alpha$,
[\ion{O}{iii}]$\lambda\lambda$4959,5007 are evident in both the MMT and
BTA spectra of HS~0837+4717. This allows us
to check the accuracy of the derived parameters.

\subsubsection{WR blue bump features}
\label{WR}

In Figure \ref{FigBroad} we show an expanded view of the BTA spectrum with
the Gaussian decomposition of the narrow lines and the broad component
of \ion{He}{ii}~$\lambda$4686 and the WR lines \ion{N}{iii}
$\lambda$4640, attributed to late WN stars,
and \ion{C}{iv}~$\lambda$4658, attributed to  WC stars (e.g., SV98).
The latter also could appear as the red bump centered at $\lambda$5808.
This bump, however, is barely detected in our spectra.
Its flux as estimated (S/N ratio $\sim$2) from the
MMT spectrum, is 2.4$\times$10$^{-16}$ erg~cm$^{-2}$~s$^{-1}$.
The parameters of fitted lines in different spectra are consistent
within their rather large uncertainties.
The parameters of narrow and broad lines are presented in Tables
\ref{t:Intens} and \ref{t:WR_broad}, respectively.

\subsubsection{The broad components of the strong emission lines}

The broad components of the strong emission lines of H$\alpha$,
H$\beta$ and
[\ion{O}{iii}]~$\lambda\lambda$4959,5007 originate in the ionized gas
moving
with velocities of $\sim$1--2 thousand \kms. Such high velocities are
related either to supernovae shells or to stellar winds of massive
stars. Since the widths of the Balmer and the [\ion{O}{iii}] lines are
similar,
it is reasonable to suggest that both originate in the same regions.
While broad features centered at H$\beta$ (the \ion{He}{ii}
$\lambda$4861
+H$\beta$ blend) and H$\alpha$ (the \ion{He}{ii} $\lambda$6562 +
H$\alpha$
blend), characteristic of WR stars (e.g., SV98),
should
contribute to the broad components of the Balmer lines, their relative
strengths calculated from models are expected to be tens of times lower
than ones observed in
HS 0837+4717. Therefore their origin in this galaxy, as well as in
other similar cases, can probably be attributed to rather young
remnants of multiple SN explosions. Having the line
intensity ratios for the broad components, we can address the physical
conditions in the regions of their formation.
The parameters of broad components of these four lines are presented in
Table \ref{t:WR_broad}.
It is evident that these components have systematically
higher line fluxes in the BTA spectra. This can be attributed to the
wider slit (2\arcsec\ at BTA vs 1\farcs5 at MMT) and to the different PA.

The Balmer decrement of the broad line components,
$I$(H$\alpha$)/$I$(H$\beta$) appears unusually large.
Accounting for the significant uncertainties,
we accept the average of the MMT and BTA ratios for the fluxes of the
broad and narrow
components of H$\beta$, equal to (2.7$\pm$0.5)\%. The same average
ratio for H$\alpha$ is (19$\pm$1.5)\%. Adopting a flux
ratio of the narrow H$\alpha$ to H$\beta$ lines, corrected for extinction,
of 2.76, a Balmer ratio for the broad components is 19.4$\pm$4.

Such a large Balmer decrement is rather unusual. To ensure its reality
we checked two possible factors, which could in principle enhance the
Balmer ratio. The first one is underlying stellar absorption in the Balmer
lines,
which has an EW comparable to the EW(broad H$\beta$), and hence could
affect it. However,
since the FWHM of stellar absorption lines ($\sim$200~\kms) is
typically $\sim$7
times smaller than that of broad emission component, the effect of
stellar
H$\beta$ absorption will decrease the flux of broad emission H$\beta$
at most by 15\%.
The second factor could be purely instrumental: light scattering in the
spectrograph of the strong lines could produce some artifacts with
arbitrary
ratio of intensities. This certainly can be excluded, since we
observed another object during the same night at the MMT and found even
stronger emission lines, but with no hints of the broad components.

Thus, this very large Balmer decrement is either related to substantial
internal extinction of the regions where the broad Balmer lines are formed,
or to the excitation of hydrogen levels by electron collisions,
or both. For the latter, H$\alpha$/H$\beta$ ratio can reach the value
of 5.8 for $T_{\rm e}$=10000~K (Chamberlain \cite{Chamber53}).
Thus, even for the case of
electron excitation, some significant extinction (accounting for the
additional factor of at least $\sim$3.3$\pm$0.7 in the Balmer line
ratio)
should be invoked for the regions of the broad Balmer lines formation.
Using
the Whitford extinction law, the latter possibility corresponds to
$E(B-V)$=1.16$\pm$0.2 and an additional extinction in the $B$-band of
$A_{\rm B}$$\sim$5$\pm$1 magnitudes! The latter indicates that the
broad line emission forms in, or propagates through, a very dusty medium,
contrary to what is observed for the narrow line \ion{H}{ii} region type
emission.
If no alternative explanation of the large Balmer decrement is found,
this implies that the fluxes of broad features, and of the unknown optical
continuum related to them, should be corrected by a factor of 25--100,
depending on the wavelength.
Such correction will raise the total luminosity of the BCG by a factor
of several.
The highly obscured regions are
quite typical of starbursts observed in mergers.
Moreover, in the XMD BCG
SBS 0335--052~E mid-IR observations reveal a highly obscured starburst
(Hunt et al. \cite{Hunt01}).

We also point out the very large difference between the ratio
of $I$($\lambda$4959)/$I$(H$\beta$) for the narrow and the broad
components:
for the average of the MMT and BTA data they are 1.88 and 0.62,
respectively.
Is this due to the effect of significant collisional de-excitation of
O$^{++}$ ions, or to a significantly lower O/H ratio in the regions of
broad
line formation, or their lower $T_{\rm e}$? It is possible that all
these factors combine to produce the observed effect.
To answer these questions,  higher S/N ratio data will be helpful.

\subsection{Spatial structure of the emission regions}
\label{structure}

Below we  exploit the advantage of long-slit spectra to
examine various morphological features of the object.
The seeing during the MMT observations ($\theta$=1\farcs6) and  BTA
observations in the red  ($\theta$=1\farcs3) was
significantly less than the full extent of both line and continuum
emission
(8\arcsec--10\arcsec), and allows the performance of a coarse analysis
of the
BCG  substructure.  Note that what we refer to as seeing is the width
of an unresolved object as measured perpendicular to the spectrograph's
dispersion
and so includes contributions from the atmosphere as well as the
telescope and spectrograph optics.

\begin{figure*}
   \centering
   \includegraphics[angle=-0,width=5.1cm]{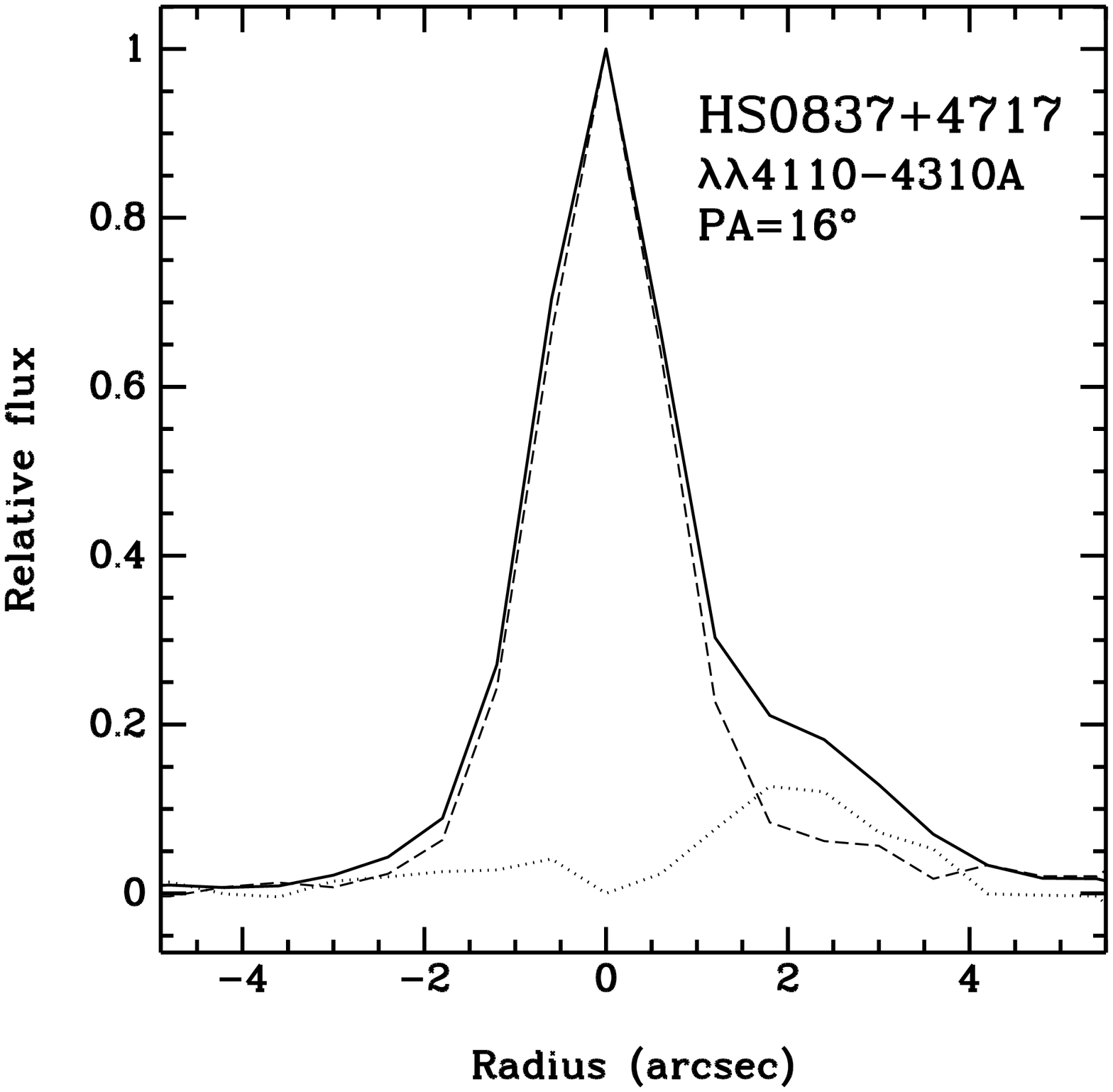}
   \includegraphics[angle=-0,width=5.1cm]{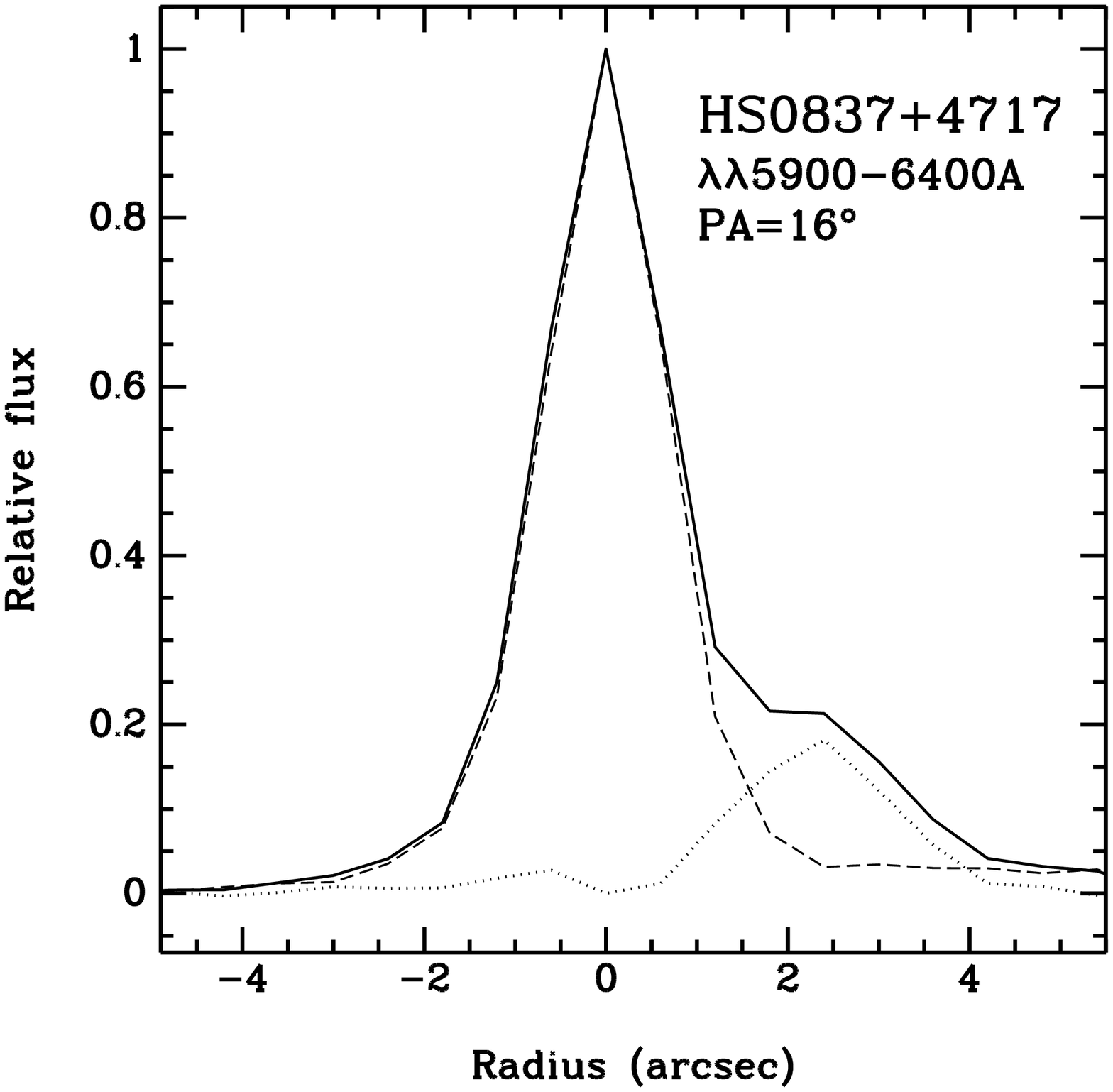}
   \includegraphics[angle=-0,width=5.1cm]{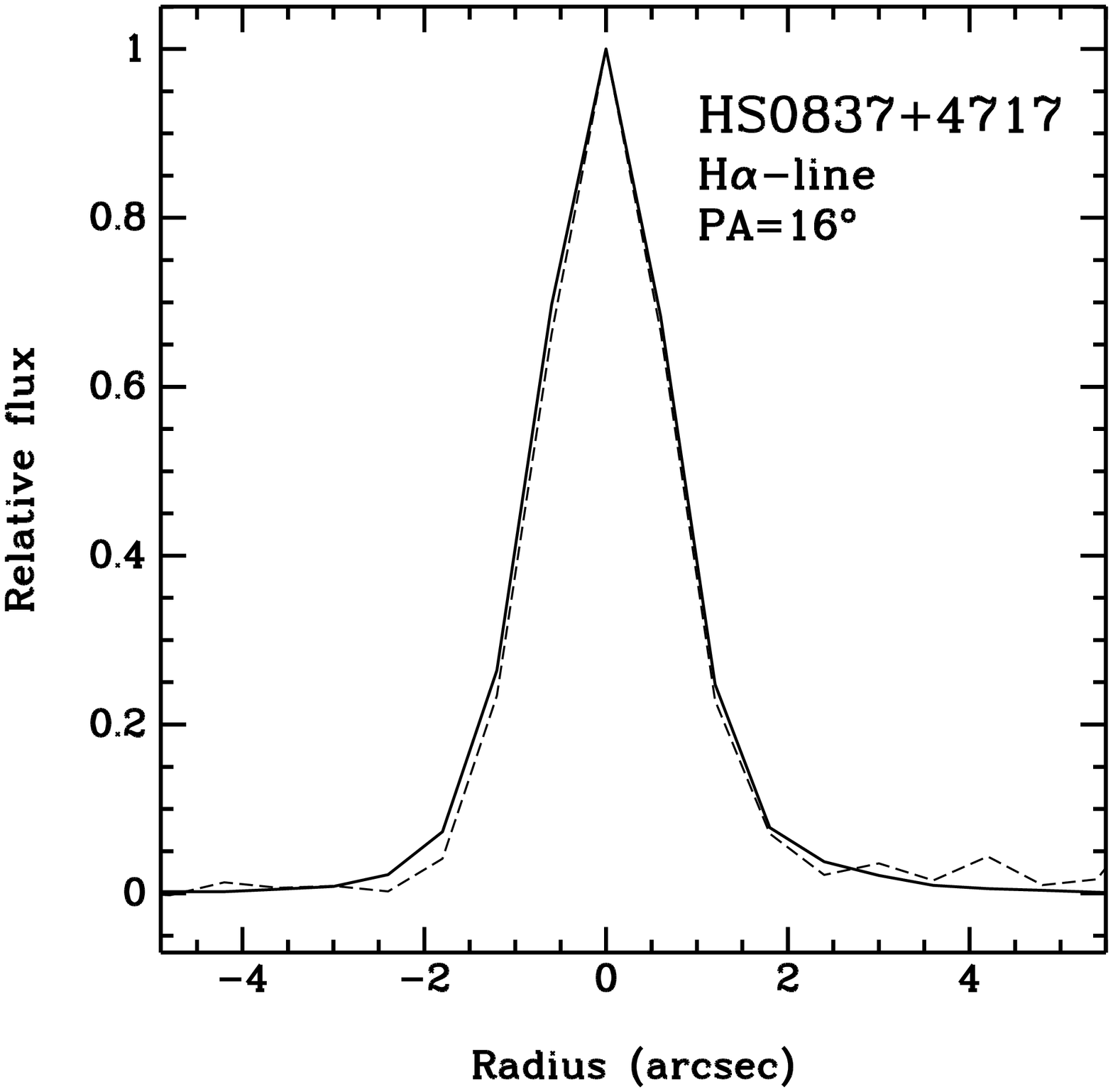}
 \caption{{\it Left panel:} Intensity distribution along the slit of the
   continuum between H$\delta$ and H$\gamma$. The dashed line
   shows the PSF (point-spread function) fitted to the central peak
   (related to the starburst). The dotted line shows the difference between
   the observed full light and that of the central bright compact region.
   The additional component, centered at $\approx$2\farcs2 to the SSW
   is $\sim$0.25 of    the bright peak amplitude.
   {\it Middle panel:} Same as in the left panel, but for continuum in
   the
   red. The relative flux of the fainter knot is 0.28, that is a factor
   of 1.13 larger than in the blue. This implies that this knot in
   $(B-R)$ is
   $\sim$ 0\fm13 redder than the central starburst, which mainly
   defines the integrated colours of the BCG.
   {\it Right panel:} Same as the left panel, but for intensity
   centered on
   the H$\alpha$ line. No significant signal is seen at the periphery
   besides the scattered light from the central bright region.
              }
         \label{cont_structure}
   \end{figure*}

In Figure \ref{cont_structure} we show the intensity profiles of the
continuum
along the slit on the MMT spectrum averaged over the line-free spectral
ranges (in the rest frame system) of 4110--4310~\AA\ and 5900--6400~\AA.
For comparison, the intensity distribution in a narrow band, centered
on the H$\alpha$+[\ion{N}{ii}] lines is also presented.

The intensity distributions of the continuum in
Fig. \ref{cont_structure} and of the strong emission lines are different.
The latter have
well fitted PSF profiles, with the FWHM close to that of seeing. Some
very faint wings/pedestal are present, which have just the appearance of
the PSF at a few per cent level of the peak value.

The spatial distribution of continuum in the BCG shows some additional
features. The main peak, related to the current starburst,
fully coincides in position (along the slit) with the
peak of the line emission. This starburst continuum component dominates the
other components. However, an additional
component (knot) on the MMT spectrum is seen at $\sim$2\farcs2
(1.8 kpc) to the SSW, with the relative flux of $\sim$25\%
in the blue
and of $\sim$28\% in the red. This implies that the $(B-R)$ of the faint
`knot' is $\sim$0\fm13 redder than that of the
current starburst, which
dominates the BCG light and determines its integrated colours. Based on
the relation above we can approximately estimate the $(B-R)$ colour of
the faint knot. From the data in Table \ref{t:Param}, the integrated
$(B-R)_{0}$ is 0.15. This colour is affected by the strong emission lines
included in both filters. Accounting for the EWs of
these lines and for the effective widths of both filters, we estimate
that the $(B-R)$ of the continuum component is redder due to the emission
lines by $\sim$0.12. This implies that the net continuum of HS 0837+4717
has a $(B-R)_{0}$ $\sim$0.03.
The latter is very similar to the colour $(B-R)$=--0.05 of the mixture
of a starburst with an age of 3.7 Myr and the continuum produced by gas at
$T_{\rm e}$=20000~K.
The former, according to the PEGASE.2 model (with a Salpeter IMF with
limits of 0.1 and 120 $M$\sunn\ and $Z$=$Z$\sunn/20), is
$(B-R)$=\mbox{--0.24}, while the latter, from the data in Aller
(\cite{Aller84}),
has $(B-R)$=+0.53. They are mixed in the proportion that results in the
observed narrow-line EW(H$\alpha$)$\sim$1400~\AA.
We accept the bright component continuum colour of $(B-R)$=--0.05, and
from the
latter we obtain for the faint knot $(B-R)_{0}^{\rm faint} \sim$0.08.
For an instantaneous star formation (SF) with the same IMF and
$Z$=$Z$\sunn/20, this
$(B-R)$ corresponds to an age of $\sim$25 Myr.
Apart from this fainter knot of continuum emission, there is also some
diffuse component, stretching further to the SW and NE, as seen
at the $V$-band image in Fig. \ref{FigDirect}.

We estimate the intrinsic size of the supergiant \ion{H}{ii} region from a
comparison of the  FWHM of its strong emission lines with that of PSF.
Unfortunately, for the sharpest images ($\theta$=1\farcs3, BTA observations in
the red), the seeing fluctuated, therefore it is difficult to
accept the value of the FWHM for the PSF. For the MMT observations the
situation is better.
A nearby star appears on the slit and can be used to estimate the PSF
at various wavelengths. The weighted mean of the FWHM(PSF)
in red and green is 1\farcs60.
The FWHM(intrinsic), derived as the weighted mean of the values in
red and green, is $\sim$0\farcs7$\pm$0\farcs07. FWHM(intrinsic)
were calculated by quadratic subtraction of the FWHM(PSF) from the measured
FWHM of strong lines.
The size of the faint knot is difficult to measure directly. However, from the
decomposition
of the light distribution along the slit, shown in Figure
\ref{cont_structure}, the FWHM of
this knot is $\sim$2\farcs2. Adopting the same FWHM of the PSF,
we obtain a coarse estimate for the intrinsic size of the faint knot as
FWHM(knot)$\sim$1\farcs5.
These sizes correspond to a FWHM linear diameters of the two components
of $\sim$0.58 kpc and $\sim$1.25 kpc, respectively.

Summarizing this analysis, we find that along the major axis at its
brightest
part, this BCG has a double-nucleus structure with characteristic sizes
of the components
of 0.6 and 1.25 kpc, respectively, and a projected distance of 2\farcs2
(1.8 kpc) between the components. The total extent of the underlying
low-surface-brightness
(LSB) component is $\gtrsim$10\arcsec\ (8 kpc). The nature of the
faint knot is not clear. But since it is also quite blue, one can
expect that this is also related to the recent SF event.

\begin{table}[hbtp]
\centering{
\caption{Abundances in the supergiant \ion{H}{ii} region}
\label{t:Chem}
\begin{tabular}{lcc} \hline  \hline
\rule{0pt}{10pt}
Value                                      & MMT                 & BTA \\ \hline
$T_{\rm e}$(OIII)(K)\                      & 18,500$\pm$700 ~~   & 18,300$\pm$800 ~~    \\
$T_{\rm e}$(OII)(K)\                       & 15,200$\pm$500 ~~   & 15,100$\pm$600 ~~    \\
$T_{\rm e}$(SIII)(K)\                      & 17,000$\pm$500 ~~   & 16,900$\pm$600 ~~    \\
$N_{\rm e}$(SII)(cm$^{-3}$)\               & 300$\pm$300~~       & 400$\pm$300~~        \\
& & \\
O$^{+}$/H$^{+}$($\times$10$^{-5}$)\        & 0.345$\pm$0.035~~   & 0.357$\pm$0.042~~    \\
O$^{++}$/H$^{+}$($\times$10$^{-5}$)\       & 3.941$\pm$0.355~~   & 3.969$\pm$0.419~~    \\
O$^{+++}$/H$^{+}$($\times$10$^{-5}$)\      & 0.082$\pm$0.014~~   & 0.095$\pm$0.022~~    \\
O/H($\times$10$^{-5}$)\                    & 4.369$\pm$0.357~~   & 4.421$\pm$0.422~~    \\
12+log(O/H)\                               & ~7.64$\pm$0.04~~    & ~7.65$\pm$0.04~~     \\
& & \\
N$^{+}$/H$^{+}$($\times$10$^{-7}$)\        & 5.157$\pm$0.701~~   & 5.265$\pm$0.660~~    \\
ICF(N)\                                    & 12.65               & 12.40                \\
log(N/O)\                                  & --0.83$\pm$0.07~~   & --0.83$\pm$0.07~~    \\
& & \\
Ne$^{++}$/H$^{+}$($\times$10$^{-5}$)\      & 0.610$\pm$0.058~~   & 0.606$\pm$0.068~~    \\
ICF(Ne)\                                   & 1.11                & 1.11                 \\
log(Ne/O)\                                 & --0.81$\pm$0.05~~   & --0.82$\pm$0.06~~    \\
& & \\
S$^{+}$/H$^{+}$($\times$10$^{-7}$)\        & 0.695$\pm$0.061~~   & 0.767$\pm$0.074~~    \\
S$^{++}$/H$^{+}$($\times$10$^{-7}$)\       & 3.437$\pm$1.294~~   & 5.196$\pm$1.282~~    \\
ICF(S)\                                    & 2.91                & 2.87                 \\
log(S/O)\                                  & --1.56$\pm$0.14~~   & --1.41$\pm$0.10~~    \\
& & \\
Cl$^{++}$/H$^{+}$($\times$10$^{-7}$)\      & 0.121$\pm$0.004~~   & ...                  \\
ICF(Cl)\                                   & 2.19                & ...                  \\
log(Cl/O)\                                 & --3.22$\pm$0.14~~   & ...                  \\
& & \\
Ar$^{++}$/H$^{+}$($\times$10$^{-7}$)\      & 1.003$\pm$0.101~~   & ...                  \\
Ar$^{+++}$/H$^{+}$($\times$10$^{-7}$)\     & 1.402$\pm$0.193~~   & 1.395$\pm$0.317~~    \\
ICF(Ar)\                                   & 1.01                & 1.71                 \\
log(Ar/O)\                                 & --2.26$\pm$0.05~~   & --2.27$\pm$0.11~~    \\
& & \\
Fe$^{++}$/H$^{+}$($\times$10$^{-7}$)\      & 0.663$\pm$0.321~~   & 0.622$\pm$0.212~~    \\
ICF(Fe)\                                   & 15.79               & 15.45                \\
log(Fe/O)\                                 & --1.62$\pm$0.21~~   & --1.66$\pm$0.15~~    \\
$[$O/Fe$]$\                                & 0.20$\pm$0.21~~     &   0.24$\pm$0.15~~    \\
& & \\
\hline   \hline
\end{tabular}
 }
\end{table}

\begin{table}[hbtp]
\centering{
\caption{Parameters of the WR lines and of the broad components}
\label{t:WR_broad}
\begin{tabular}{lrr} \hline \hline
\rule{0pt}{10pt}
& MMT & BTA    \\ \hline
\MC{3}{c}{Flux (10$^{-16}$ ergs\ s$^{-1}$cm$^{-2}$)$^a$}    \\
4640\ N\  {\sc iii}\ (WR)\  &   2.5 &   3.8 \\
4658\ C\  {\sc iv}\ (WR)\   &   2.6 &   5.0 \\
4686\ He\ {\sc ii}\  (WR)\  &   2.4 &   3.7 \\
4861\ H$\beta$\             &   6.3 &  15.0 \\
4959\ [O\ {\sc iii}]\       &  11.6 &  18.3 \\
5007\ [O\ {\sc iii}]\       &  34.8 &  55.0 \\
6563\ H$\alpha$\            & 159.4 & 261.3 \\  \hline
\MC{3}{c}{Equivalent width [\AA]}                         \\
4640\ N\  {\sc iii}\ (WR)\  &   1.4 &   2.1 \\
4658\ C\  {\sc iv}\ (WR)\   &   1.5 &   2.8 \\
4686\ He\ {\sc ii}\  (WR)\  &   1.4 &   2.1 \\
4861\ H$\beta$\             &   4.0 &   7.8 \\
4959\ [O\ {\sc iii}]\       &   7.9 &   9.8 \\
5007\ [O\ {\sc iii}]\       &  24.3 &  30.0 \\
6563\ H$\alpha$\            & 238.4 & 302.2 \\  \hline
\MC{3}{c}{FWHM [km s$^{-1}$]}                         \\
4640\ N\  {\sc iii}\ (WR)\  & 1300 &  1300 \\
4658\ C\  {\sc iv}\ (WR)\   & 1300 &  1300 \\
4686\ He\ {\sc ii}\  (WR)\  & 1300 &  1300 \\
4861\ H$\beta$\             & 1500 &  1500 \\
4959\ [O\ {\sc iii}]\       & 1500 &  1500 \\
5007\ [O\ {\sc iii}]\       & 1500 &  1500 \\
6563\ H$\alpha$\            & 1300 &   900 \\  \hline
\MC{3}{c}{Broad/Nebular flux ratio [\%]}               \\
4686\ He\ {\sc ii}\  (WR)\  & 36  &  46 \\
4861\ H$\beta$\             & 1.9 & 3.5  \\
4959\ [O\ {\sc iii}]\       & 1.7 & 2.2  \\
5007\ [O\ {\sc iii}]\       & 1.7 & 2.2  \\
6563\ H$\alpha$\            & 16  &  22 \\  \hline
\MC{3}{c}{WR/O stars ratio}                      \\
Observed                    & \MC{2}{c}{0.02~~~~~~~~~~~~~~~~~~~~}  \\
Model$^b$                   & \MC{2}{c}{0.02~~~~~~~~~~~~~~~~~~~~} \\
\hline \hline
\MC{3}{l}{$^a$ Flux errors are $\sim$50\% for the broad \ion{He}{ii}~$\lambda$4686 line and}\\
\MC{3}{l}{~~$\sim$(15--25)\% for the broad components of the strong lines.}\\
\MC{3}{l}{$^b$ data from Schaerer \& Vacca (\cite{SV98}).}\\
\end{tabular}
 }
\end{table}

\subsection{Ionized gas kinematics}

\begin{figure*}
   \centering
   \includegraphics[angle=-0,width=7cm]{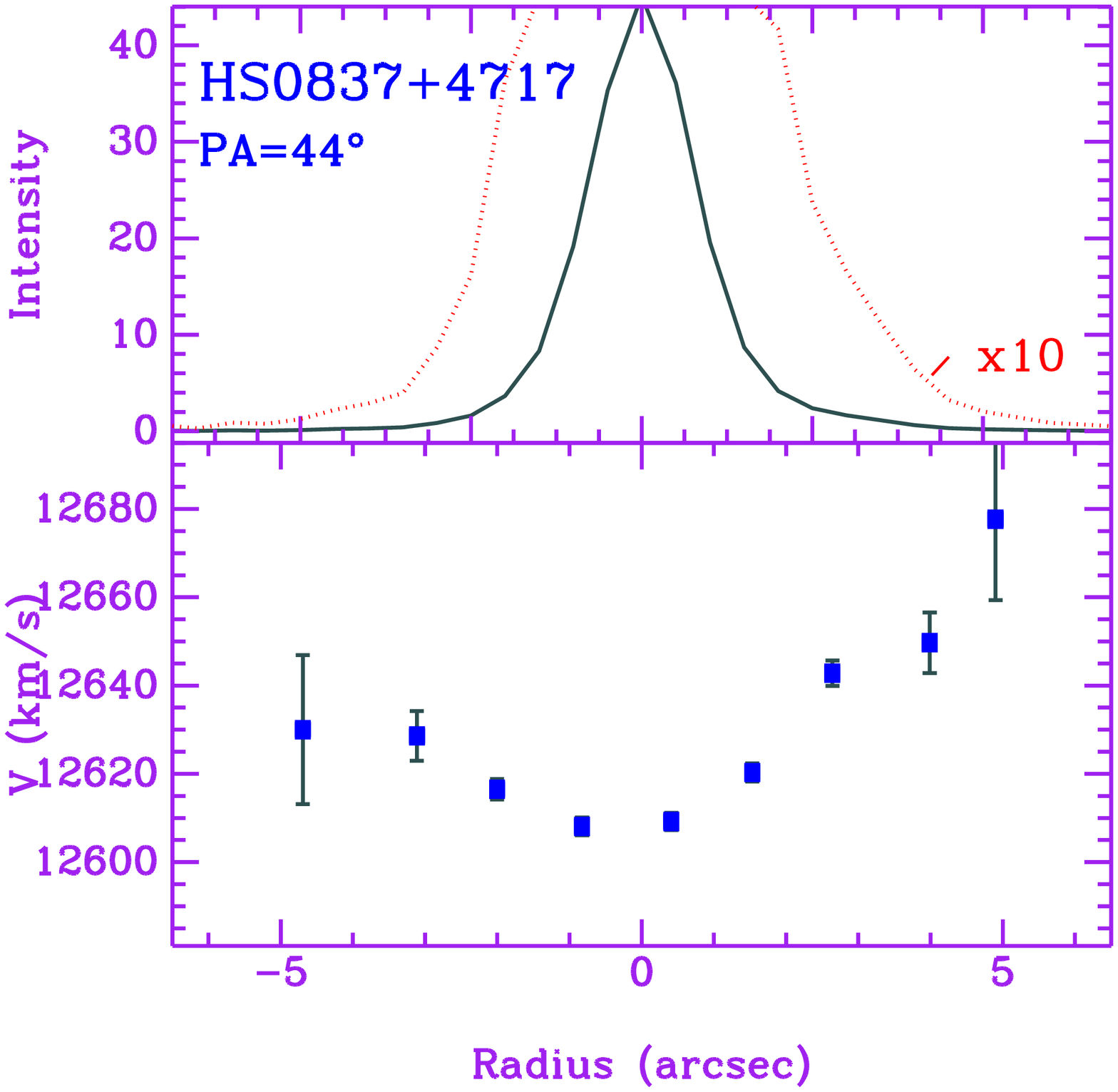}
   \includegraphics[angle=-0,width=7cm]{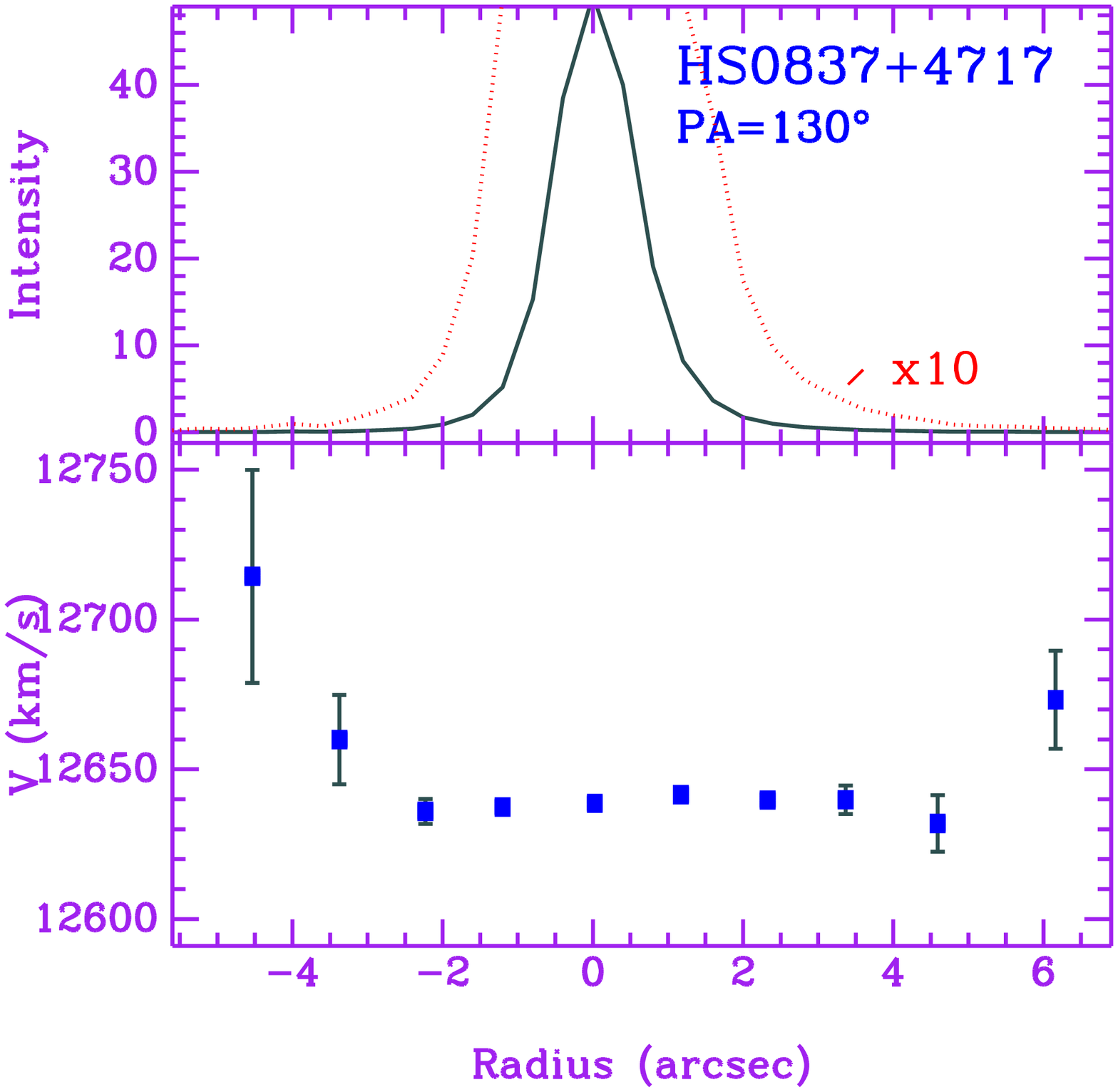}
      \caption{{\it Left panel:} Intensity distribution of H$\alpha$
       along the slit at
       PA=44\degr\ and the respective P--V diagram (binned
       with 3 pixels, or 1\farcs2, corresponding to the seeing).
       The radial velocity increases monotonously
       in the SW direction, resembling
       the receding
       branch of rotation curve with the amplitude of 50--70 \kms.
       However, the expected approaching branch for negative
       coordinates
       in the direction to NE is absent (see text for discussion).
       {\it Right panel:} Same as in the left panel but at PA=130\degr.
      The velocity variations in the central part of this cross-section
      are significantly smaller than those at PA=44\degr. This would be
      expected due to the small projection of the rotation velocity onto the
      minor axis. There seems to be a large velocity disturbance  at the NW
      edge, where radial velocity steeply raises to 60--70 ($\pm30$)
      \kms\ in the two last points. The small velocity variations are
      also observed on the opposite side of the galaxy.
              }
         \label{ROT44}
   \end{figure*}

We performed an analysis of the gas kinematics following the method
described by Zasov et al. (\cite{Zasov00}).
The H$\alpha$ line position-velocity (P--V) diagrams
along the slit at PA=44\degr\ and 130\degr\ are shown in Fig.
\ref{ROT44}. The first one is close to the
major axis of the outermost isophotes, while the second is nearly
perpendicular.
The full extent of H$\alpha$ emission as seen from these spectra is
the same in both directions: 9\arcsec--10\arcsec\
(or $\sim$7--8 kpc).  The appearance of \mbox{P--V} diagrams, however,
is very different. For PA=130\degr\  the radial velocity is practically
constant for a range from --3\arcsec\ to $\sim$3\farcs5, or about 65\%
of the entire extent of the observed H$\alpha$-emission. This is
consistent with a negligible expected  projection of the rotation
velocity for this PA.
The velocity curve at the galaxy edges looks, however, somewhat
disturbed. This could be
a result of a recent interaction or some indication for a gas blowout,
confined in some directions. Better S/N ratio data for these regions
are necessary to confirm the significance of the velocity disturbances.

The slit position at PA=44\degr\ is expected to be the optimal one to
see the overall rotation of this galaxy.
However, the respective P--V diagram does not look like a typical
rotation.
We discuss below two possible explanations for the unusual distribution
of the radial velocity.

The first one treats the P--V diagram as the consequence of
counter-rotation.
At positive coordinates along the slit we see the receding
branch of the rotation curve with a velocity
amplitude of 50--70 \kms\ (depending on how the real curve is drawn
between
the two last points) relative to the brightness peak velocity.
However, the symmetric (approaching) continuation of this branch at
negative
coordinates is not seen. Instead, the receding branch is present,
with an amplitude of 20--30 \kms. This could probably indicate 2
dynamically decoupled gas systems.
A difference of $\sim$30~\kms\ between the radial velocities near the peak
of the H$\alpha$ emission for PA=130\degr\ and
PA=44\degr\ can be attributed to small difference in the zero-points
of the dispersion curves for the two independent spectra.

Another explanation treats the part of P--V diagram at negative coordinates
as a significant velocity disturbance relative to the extrapolated
approaching
branch. This could be related to a strong outflow from the recent SF
episode.
If one is to draw the approaching branch of the rotation curve as a
symmetric extrapolation of the receding one, the resulting difference
between
the observed data and drawn approaching branch shows us the form of the
disturbance. This resembles the common wave-like form of supershells,
as they are usually seen on P--V diagrams of starbursting galaxies.
The velocity of such supershell relative to the background rotation
velocity is $\sim$70--100~\kms. Its visible extent is $\sim$2\arcsec.
The results could indicate that only a half of the shell is seen, which
is closer to the center.
Therefore, its full extent could be $\sim$4\arcsec, or $\sim$3.3 kpc.
Such features in \mbox{P--V} diagrams are detected in many BCGs, and
the most
plausible explanation is related to the expansion of supershells formed
in the ionized gas
due to the recent star-forming activity in these objects.
Supershells are produced by hot bubbles, originating from the
energy injection of numerous massive star winds and supernovae (SNe)
explosions into the galactic ISM (e.g., Tenorio-Tagle \& Bodenheimer
\cite{TTB88}).

In particular, a similar, but scaled down according to the parameters
of the host galaxy, supershell in HS 0822+3542 was discussed
by Pustilnik et al. (\cite{SAO0822}). The energy of that supershell
with a diameter of 0.48 kpc and a velocity amplitude of 30 \kms\ was
provided by 13 SNe. Assuming the same density of ambient gas of 0.1
cm$^{-2}$,
with the supershell parameters as observed in HS 0837+4717 (adopting a
minimal $V_{\rm shell}$=70~\kms), we derive a total
energy larger by a factor of $\sim$1770. This corresponds to $\sim$23000
SNe exploded during a period of the order of one dynamical age of the
supershell. The latter is
$t_{\rm dyn}$ = 0.6 $R_{\rm shell}/V_{\rm shell}$=13.7 Myr
(Weaver et al. \cite{Weaver77}). This would imply the existence of a
star cluster with an age of $\sim$17 Myr, provided the mean rate $\sim$1700
SNe per Myr during last $\sim$14 Myr (recall, that SNe begin to explode
in the cluster after $\sim$
3 Myr). From this SN rate and a PEGASE.2 model with the same IMF
as above, we derive the mass of such a cluster of
$\sim$7.6$\times$10$^{6}~M$\sunn.

For a galaxy of this high luminosity and an apparent diameter of
$\sim$10 kpc, the estimated rotation velocity is rather small.
The inclination correction, if applicable, is modest.
From the apparent axial ratio of this BCG ($p=b/a=0.67$), it
follows, that in the case of a stable rotating disk, $i$ is quite large.
Indeed, from the well known formula:
$cos^{2}(i)=(p^{2}-q^{2})/(1-q^{2})$, and the range of
intrinsic ratio for disk-like galaxies of $q=b/a=0.2-0.3$, the $i$
would fall in the range of 49\degr--52\degr.
The respective correction of 1/$\sin$($i$)=1.27--1.32.
In this case the corrected $V_{\rm rot}$ can reach the value of 65 to 90
\kms, within the range characteristic  of low-mass galaxies.

If this object is a merger, gas motions due
to dissipative collisions could be significantly decoupled from the
motion of stars.
Until the system relaxes, the gas motions are bad tracers of mass
distribution (e.g., Amram \& \"Ostlin \cite{AO01}). Therefore the amplitude
of the rotation velocity observed in H$\alpha$ could be unrelated
to the overall gravitational potential.

\subsection{Imaging and photometry}
\label{imaging}

In the light of the discussion on the possible merger nature of
HS~0837+4717,
it is important to inspect the 2D morphology of the galaxy.
Its grey-scale $V$-band image with isophotes superimposed is shown in
Figure \ref{FigDirect}.

The overall structure is asymmetric and somewhat disturbed,
both near the center, and in the peripheral regions. In particular, the
fainter component at $\sim$2\arcsec\ to the South (which is seen in the
continuum on the long-slit spectrum) is apparent on this image, however
smeared due to the seeing effect. The outermost isophotes indicate
several appendages to the main body.
Disregarding the faintest isophote as being subject to noise,
we notice two small features,
at NNW and SSE of the center, the latter being much more prominent.
The BCG images
available from the Data Release 1 (DR1)
database
of the Sloan Digital Sky Survey (SDSS) (York et al. \cite{York2000},
Abazajian et al. \cite{DR1}) corroborate this disturbed morphology.

As described above, small transparency variations took place during
imaging with the Wise 1-m telescope.
In order to minimize their effect on the object's integrated magnitudes,
we selected from each of the nights the brightest value in each of the
bands, and compared them with the magnitudes, derived through the
convolution of the MMT spectrum. For $U$-band we could not use an MMT
measurement due to the limited wavelength coverage. Therefore, even
though the $U$-band magnitudes for both nights are consistent within the
observational
uncertainties, they should be treated with caution. The brightest
values
for the Wise measurements in $B$ and $R$ are consistent with the MMT
data within the uncertainties, therefore we accept
them with an r.m.s. error of 0\fm10.  In the $V$-band, however, the
brightest magnitude on Wise data is still 0\fm27 fainter than the
respective
magnitude derived from the MMT spectrum. Therefore, for $V$ we accept
the MMT derived value. The integrated photometry data are summarized
in Table~\ref{t:Param}.

The very red $(B-V)$ ($\sim$0.9) and blue $(V-R)$ ($\sim$--0.7) colours of
this object are related to the very large EWs of the [\ion{O}{iii}]
$\lambda$4959 and $\lambda$5007~\AA\ emission lines, redshifted to the
center of $V$-band. This can be checked by the convolution of the
deredshifted BCG spectrum (see Table \ref{t:Param}).

\subsection{\ion{H}{i} results}

Effectively, we observed HS~0837+4717 in "on+off" positions for 320
minutes and, after smoothing, obtained an r.m.s. noise of 1.3 mK. This
translates to an r.m.s. in flux density of 1.6 mJy. Since no signal
was detected, we put  a 2$\sigma$ upper limit of 3.2 mJy on the peak
flux density from the galaxy.
The full amplitude of ionized gas velocity variations, $\sim$100--140
\kms\
is used to estimate the upper limit of the \ion{H}{i} flux. Adopting
this
value as an estimate of $W_{\rm 50}$ for the \ion{H}{i} profile,
the
upper limit of the integrated \ion{H}{i} flux is of (0.32--0.45) Jy~\kms.
The respective upper limit on the \ion{H}{i} mass is $M$(\ion{H}{i},BCG)
$\le$ (2.1--3.0)$\times$10$^{9}~M$\sunn.

\section{Discussion and conclusions}
\label{discussion}

\subsection{General comments}

The integrated parameters of HS~0837+4717 are summarized in Table
\ref{t:Param}.
There are two unusual properties, which are probably related. The first
one is the high luminosity of this BCG for its very low metallicity (see,
e.g., plots of O/H versus $M_{\rm B}$ in Kunth \& \"Ostlin \cite{Kunth2000};
Pustilnik et al. \cite{Kiel03}).
The second one is the very large nitrogen abundance excess: N/O is a
factor of $\sim$6  larger than for other BCGs with similar O/H.
The first property can be attributed to an exceptionally strong
starburst, which
causes an additional brightening by several magnitudes. Pustilnik
et al. (\cite{PKLU}) argued that the majority of starbursts in BCGs are
triggered by interactions of various strengths with other galaxies. The
range spreads from weak tidal interactions through strong tidals and
the sinking of satellites to
merging of low-mass galaxies. The latter case can be considered as a
down-scaled analog of ultraluminous IR galaxies, mergers of
massive gas-rich galaxies (e.g., Genzel et al. \cite{Genzel01}).
Below we discuss the possibility that HS 0837+4717 as well could be
an advanced merger.


\begin{table}
\caption{Main parameters of HS~0837+4717}
\label{t:Param}
\begin{tabular}{lc} \\ \hline \hline
Parameter                            & Value                \\ \hline
R.A.(J2000.0)                        & ~~08 40 29.84        \\
DEC.(J2000.0)                        & $+$47 01 09.1        \\
$A_{\rm B}$$^{(N)}$                  & 0.11                 \\
$B^{\rm tot}$                        & 18.18$\pm$0.10$^{(2)}$       \\
$(U-B)^{\rm tot}_0$                  &--0.41$\pm$0.14$^{(2)}$~~    \\
$(B-V)^{\rm tot}_0$                  &~~0.86$\pm$0.14$^{(1,2)}$    \\
$(V-R)^{\rm tot}_0$                  &--0.71$\pm$0.14$^{(1,2)}$    \\
V$_{\rm Hel}$ (\kms)                 & 12630$^{(2)}$           \\
Distance (Mpc)                       & 170.5$^{(2)}$           \\
$M_{\rm B}^0$$^{(3)}$                &  --18.1$^{(2)}$             \\
Opt. size (\arcsec)$^{4}$            & $\sim$8$\times$7$^{(2)}$     \\
Opt. size (kpc)                      & 6.5$\times$6$^{(2)}$         \\
12+$\log$(O/H)  \                    & 7.64$\pm$0.03$^{(2)}$  \\
$T_{\rm e}$(\ion{O}{iii}) (K)\       & 18000$^{(2)}$         \\
$N_{\rm e}$  (cm$^{-3}$)\            & $\lesssim$400$^{(2)}$  \\
log(N/O)\                            & --0.83$\pm$0.05$^{(2)}$\\
\ion{H}{i} int.flux$^{(5)}$          & $<$0.31--0.43$^{(2)}$        \\
$M$(\ion{H}{i}) (10$^{9}~M$\sunn)    & $<$2.5$^{(2)}$     \\
$M$(\ion{H}{i})/$L_{\rm B}$$^{(6)}$  & $<$0.9$^{(2)}$    \\   \hline\hline

\multicolumn{2}{l}{(1) -- ($B-V$)$_{\rm tot}^0$ corrected for redshift is 0\fm21.} \\
\multicolumn{2}{l}{(1) -- ($V-R$)$_{\rm tot}^0$ corrected for redshift is --0\fm02.} \\
\multicolumn{2}{l}{(2) -- this paper.} \\
\multicolumn{2}{l}{(3) -- corrected only for the Galactic extinction.} \\
\multicolumn{2}{l}{(4) -- $a \times b$ at $\mu_{\rm B} =$25 mag/$\Box$\arcsec.} \\
\multicolumn{2}{l}{(5) -- for $W_{\rm 50} \le$100--140~\kms, in Jy$\cdot$\kms.} \\
\multicolumn{2}{l}{(6) -- in units ($M/L_{\rm B}$)\sunn} \\
\multicolumn{2}{l}{($N$) -- data from the NED.} \\
\end{tabular}
\end{table}

\subsection{Possible evidences of merger}
\label{morphology}

Telles and Terlevich (\cite{TT97})
found
that the most luminous BCGs show disturbed external morphologies, while
low luminosity BCGs show regular morphology.
In their detailed studies of
several luminous BCGs, \"Ostlin et al. (\cite{Ostlin99,Ostlin01})
found morphological and kinematic evidences for the
merger nature of these galaxies. HS~0837+4717 is similar to their BCGs
in several aspects.

As shown above, the overall disturbed morphology of this BCG, its
"double-nucleus" appearance,
and the presence of
appendages on the periphery, presumably imply some recent disturbance.
However, no disturbing galaxies are visible in the
environment of this BCG (see Sect. \ref{Global}).

In addition, the distribution of the ionized gas velocity also appears
disturbed, mainly near the edges of the galaxy.
The gas motions along the
apparent major axis show large deviations from the expected regular
rotation.
These could be related either to a counter-rotating gaseous component,
or to a very energetic supershell.
The two knots, related to the current/recent starbursts
are very massive.
Indeed, from the photometry in table \ref{t:Param} and numbers derived
in Sect. \ref{structure}, from a PEGASE.2 model with $Z$=$Z$\sunn/20,
and a Salpeter IMF as above, we find the following.
For the brighter component with $M_{\rm B}$=--17.9 and an age of 3.7 Myr
the mass of starburst is 0.8$\times$10$^{7}~M$\sunn.
As shown above, a significant part of this starburst could be highly
obscured by dust. Therefore, its unobscured blue luminosity could be
several times higher than that listed here. This would translate to a
total mass for this starburst region of $\gtrsim$3$\times$10$^{7}~M$\sunn.
For the faint knot $M_{\rm B,faint}$=--16.4.
Its colour $(B-R)$ corresponds to an instantaneous SF burst with an age
of 25 Myr.
With the same IMF as above, PEGASE.2 predicts
a star
cluster mass of 2.7$\times$10$^{7}~M$\sunn. Also, from the parameters
of the suggested supershell at the NE edge of the galaxy, we estimate
the mass of the embedded star cluster with an age of $\sim$17 Myr as
$\sim$0.76$\times$10$^{7}~M$\sunn. Such massive starbursts should
involve
molecular clouds with the masses of 10$^{8}~M$\sunn\ or more. We will
return to this point in Sect. \ref{Phenomenon}.
Finally, as shown, e.g., by Mihos \& Hernquist (\cite{Mihos96}), very
energetic starbursts are characteristic  of merging galaxies.

An alternative option for such a case is a strong interaction with
an undetected massive object, such as  very LSB
galaxy or a hydrogen cloud devoid of stars, similar to the case of
Dw 1225+0152 (Salzer et al. \cite{Salzer91}).

\begin{table*}
\caption{Starburst galaxies with significant nitrogen excess}
\label{t:group}
\begin{tabular}{llrllcll} \\ \hline \hline
IAU Name    &Synonyms& $V_{hel}$$^a$& $B_{tot}$$^a$ &$M_{B}$ &12+$\log$(O/H)& $\Delta \log$(N/O)&   Notes \\ \hline  
0034$-$3349 & Haro11 &  6175    & 14.6      &--20.0  &  7.90        &     0.8           &   merger                          \\ 
0125$-$061  & MKN996 &  1622    & 15.1      &--17.1  &  8.00        &     0.6--1.4      &   probable merger                 \\ 
0218$+$003  & UM420  & 17514    & 16.5      &--20.3  &  7.89        &     0.5           &   sinking merger?                 \\ 
0459$-$043  & MKN1089&  4107    & 15.0(13.3)&--19.1(--20.8)&  8.07  &     0.5           &   bright knot in interacting gal. \\ 
0837$+$4717 & HS,PC  & 12630    & 18.2      &--18.1  &  7.64        &     0.77          &   merger?                         \\ 
1139$+$006  & UM448  &  5560    & 14.7      &--19.9  &  7.98        &     0.5           &   2 knots in center -- merger?    \\ 
1337$-$313  & NGC5253&   440    & 10.9      &--17.4  &  8.16        &     0.5           &   Im pec, starburst               \\  \hline \hline 
\multicolumn{8}{l}{$^a$ data from the NED.} \\
\end{tabular}
\end{table*}

\subsection{WR star population}
\label{WR_number}

The WR stellar population is observed only in four of
about
30 XMD BCGs known. They include I~Zw~18 (Izotov et al. \cite{Izotov97a};
Legrand et al. \cite{Legrand97}), SBS~0335--052E (Izotov et al.
\cite{Izotov99}), Tol~1214--277 (Fricke et al. \cite{Fricke2001}) and
UM 133 (Kniazev et al.
\cite{3_UM}, 2004, in prep.). Thus, HS 0837+4717 is the fifth XMD BCG with a
detected
WR population. While the accuracy of the flux measurements of the broad
WR features on the available spectra is far from sufficient for accurate
prediction, it is useful to
obtain an estimate of the WR-to-O star relative numbers.

To derive $N$(WR)/$N$(O+WR), we follow the approach
described in detail by Guseva et al. (\cite{Guseva2000}).
The number of WR stars was derived from the luminosity of the WR lines.
Their fluxes were derived directly from the 1-D spectra,
as discussed in Sect. \ref{WR}.
No aperture correction was applied to the WR-line fluxes, assuming that
this emission
is produced in a compact region with an angular size less than the
width of the slit.
The number of O stars was derived from the total H$\beta$ luminosity
for the brightest SF region ($L_{\rm cor}$),
corrected for the contribution of WR stars and for the
aperture effect. For the latter we adopt the circular symmetry of
the bright SF region as seen in the direct image.
To estimate the fraction of the undetected H$\beta$ flux,
the radial distribution of the H$\beta$ line intensity,
shown in Figure \ref{Fig1DSpectra}, was used.
The respective correction factor for the total H$\beta$ flux is 1.66.

The number of O stars is derived from the number of ionizing photons
$Q_0^{\rm cor}$, which is related to the total luminosity
of the H$\beta$ emission line $L_{\rm cor}$(H$\beta$) by
\begin{equation}
L_{\rm cor}({\rm H}\beta) = 4.76\times10^{-13}Q^{\rm cor}_0.
\end{equation}

For a representative O7V star we adopt the
number of Lyman continuum photons emitted to be
$Q_0^{{\rm O7V}}$ = 1 $\times$ 10$^{49}$ s$^{-1}$ (Leitherer
\cite{L90}).
The total number of O stars is then derived from the number of O7V
stars by
correcting for other O star subtypes, using the parameter $\eta_0$
introduced by Vacca \& Conti (\cite{V92}) and Vacca (\cite{V94}).
$\eta_0$ was calculated in SV98
as a function of the time elapsed since the onset of an instantaneous
star
formation burst, as inferred from the
$EW$(H$\beta$).
The Salpeter IMF with the $M_{\rm low}$ and $M_{\rm up}$
limits of 0.8 $M_\odot$ and 120 $M_\odot$ was adopted here. We assumed
that the average Lyman continuum flux per one WR star $Q^{\rm WR}_0$
is comparable to $Q_0^{\rm O7V}$ and is equal to
1.0 $\times$ $10^{49}$ s$^{-1}$ (Schaerer et al. \cite{S99}). Finally
\begin{equation}
N({\rm O})=\frac{Q^{\rm cor}_0-N_{\rm WR}Q^{\rm WR}_0}{\eta_0(t)Q^{\rm O7V}_0}.
\label{eq:NO}
\end{equation}

In Sect. \ref{WR} we performed the estimates of
the observed fluxes of all broad WR lines in the blue and red bumps
(see Table \ref{t:WR_broad}).
The latter are used to get
the luminosities in the respective lines.
Then, the luminosity of red bump is used to estimate the number of
representative WC4 stars, having $L_{5808}$=3.0$\times$10$^{36}$ erg
s$^{-1}$: $N_{\rm WC}$=280.
The number of representative WN7 stars can be estimated either from
its adopted luminosity in broad \ion{He}{ii} line, or from the sum luminosity
of the blue bump, subtracting the contribution of \ion{C}{iv}$\lambda$4658,
caused by WC stars. The uncertainties of all WR broad features are large,
about 50\%. Therefore the derived numbers of WR stars are only indicative.
Better signal-to-noise spectra are needed to study the WR population in this
BCG at the quantitative level.

Assuming the luminosity of a single WN7  star to be 1.6$\times$10$^{36}$
ergs s$^{-1}$ in the broad \ion{He}{ii} $\lambda$4686 line or
1.99$\times$10$^{36}$ ergs s$^{-1}$ in the blue bump (SV98), we estimate
from the WR line
fluxes derived on
the MMT spectrum,
$N_{\rm WN}$ to be either 500, or 860. Adopting the average
of the two estimates for WN stars, we get the sum number of WR stars
(WN+WC) in the young starburst to be $\sim$1000.

Following to SV98
and using the observed value of $EW$(H$\beta$) = 230~\AA,
we estimate the burst age and the parameter $\eta_0$  (updated model
version of 1999, as presented in Starburst99) in
HS 0837+4717 as 3.7 Myr, and $\sim$0.65, respectively.
The total number of O stars in HS~0837+4717 is found
to be N(O) $\approx$ 47000. This yields a relative number
of WR stars to O stars \mbox{$N$(WR)/($N$(O + WR)} of 0.02,
coinciding with the model prediction in SV98.

\subsection{A large excess of the nitrogen abundance}
\label{Phenomenon}

The possibility of enhanced N abundance in galaxies with the spectral
signatures of WR stars, in comparison to other galaxies of similar
metallicity
was first noticed by Pagel et al. (\cite{Pagel86}) where no details on
the strength of this effect were given. The occurrences of a {\it
significant}
nitrogen excess  in galaxies  are very rare. For example, among more
than one hundred BCGs with well measured heavy element
abundances (e.g., IT99; Ugryumov et al. \cite{HSS-LM})
only four galaxies (Mkn 996, UM 420, Mkn 1089 and UM 448)
exhibit a  nitrogen excess of a factor of 3 and larger,
relative to the value appropriate for their O/H (Thuan et al.
\cite{Thuan96}; IT99).
All four have 12+$\log$(O/H)$\gtrsim$7.9  and show broad WR
lines (Guseva et al. \cite{Guseva2000}; Thuan et al. \cite{Thuan96}).
One more similar galaxy, Haro 11 is found among the luminous BCGs studied by
Bergvall \& \"Ostlin (\cite{Bergvall02}).
With
12+$\log$(O/H)=7.9
it has
a nitrogen abundance excess
of a factor of $\sim$6. Strong signatures of WR population are detected
in this BCG as well.

An excess of the nitrogen abundance by a factor of 3 is also detected
in two central \ion{H}{ii} regions (both with detectable WR population)
of the nearby starbursting dwarf irregular galaxy NGC~5253
with
12+$\log$(O/H)=8.16 (Kobulnicky et al. \cite{Kobul97}, and references
therein). In Table \ref{t:group} we summarize the parameters of
all known such
galaxies,
and below
draw some tentative  empirical correlations.
First, most of these galaxies could be identified as various
stages of mergers. Second, as a subgroup of starbursting galaxies,
70\% of them are luminous
($M_{\rm B}^{0}<$--18\fm0), thus are not classical dwarf
galaxies.
Third, all these
objects are WR
galaxies.
Fourth, all these BCGs but one have 12+$\log$(O/H) in
the range of $\sim$7.9 to 8.16. HS 0837+4717 is the only such galaxy
belonging to the XMD group.

The scenarios of localized nitrogen enrichment, as well as the
constraints
imposed by the known mechanisms of fresh elements mixing and dispersal
were
discussed in detail by Kobulnicky et al. (\cite{Kobul97}) (see also Roy
\& Kunth \cite{Roy95}). They concluded that, in the light of the expected
short timescales for dispersal in the interstellar medium (ISM) of a few Myr,
the localized nature of the N enrichment suggests a very recent pollution
event, probably connected with the onset of massive star winds.
The alternative analysis of Tenorio-Tagle (\cite{TT96}) suggests the
timescale
of $\sim$100 Myr to mix fresh metals from a cluster of massive stars with
the gas in a disk galaxy. However, this can be barely applicable for very
complex flow patterns expected in the vicinity of merger starbursts.
In particular,
the very
rare occurrence of objects with a large N overabundance, similar to that
observed in Mkn 996 and HS 0837+4717,
favors
the general idea of
the both short-time scales: for localized N pollution and its fast
dispersal.

It is worth noting that a selection effect,
related to the starburst strength and its trigger mechanism,
will
affect the observed frequency of this phenomenon,
probably determining the overall scale of the polluted region.
In particular,
the large N excess
in the
central starbursts of NGC 5253
would be barely detected with ground-based spectroscopy, if this galaxy
was a few
times more distant.

\subsection{The global environment}
\label{Global}

On the large scales, the volume
around HS 0837+4717
is well sampled only by the objects from the Updated Zwicky Catalog
(Falco et al. \cite{Falco99}), which is complete to $B =$15\fm5.
At the BCG distance
they all are luminous,
with $M_{\rm B} \lesssim$\mbox{--20\fm6},
and are located
outside the sphere with $R$ = 8.4 Mpc, centered at HS 0837+4717.
This implies that HS 0837+4717 is situated in a void
(as several other XMD BCGs are, Pustilnik et al. \cite{SAO0822}).
New data on the distribution of fainter galaxies with measured redshifts
from the SDSS DR1 database (Abazajian et al. \cite{DR1}) around HS 0837+4717
show no objects within this sphere, corroborating the classification
HS 0837+4717  as a void galaxy.


\subsection{HS~0837+4717 in relation to other XMD BCGs}

One of the goals of detailed studies of the properties of individual
XMD BCGs is to establish how homogeneous is this group of galaxies.
Current models suggest various evolution scenarios that would result
in a large metal deficiency of a galaxy's ISM. They include both a
significant loss of fresh metals via galactic superwinds, related to
strong starbursts, and/or the infall of unprocessed intergalactic matter
onto normally evolved galaxy. The third option is a very slow astration
due
to very low surface gas density and the stabilizing role of the Dark
Matter
(DM) halos (as, e.g., for LSB galaxies). Finally, one of the most
intriguing options is that of truly young XMD galaxies.

Known XMD galaxies can, in principle, be the products of any of these
evolutionary scenarios. Therefore,
their properties could show significant diversity. HS~0837+4717 is, in
this context, an outlier among XMD BCGs and could yield important
information on alternative evolutionary scenarios.

What are the main differences between HS 0837+4717 and
prototypical
XMD BCGs
I~Zw~18 and SBS~0335--052E,
both known to show WR features  (Sect. \ref{WR_number}),
but not an excess nitrogen abundance?
The BCG  discussed here, is the most luminous
and is probably the largest galaxy in
this
group. The latter
implies a rather massive baryon and DM configuration, capable
in the isolated state
of
maintaining a large fraction of the newly produced metals.
Its
value of $M$(\ion{H}{i})/$L_{\rm B} \lesssim$0.9
implies
either significant processing of neutral gas,
or a very strong starburst,
or both.
Also, a significant gas mass fraction could be
hidden in molecular form. The latter is more
characteristic of
luminous IR
galaxies, a class of galaxies tightly connected to powerful starbursts
due to the merging of gas-rich galaxies.
The strong starbursts, in turn,
occur
when they encompass
a significant part of the galaxy gas mass,  what
is expected in case of
strong disturbances.  Merging of two galaxies
provides the strongest possible
disturbance. This allows the breaking of the gas equilibrium even in
very stable configurations, and results in the efficient gas sinking
into
the dynamical center of merged object with a subsequent intense SF
burst (e.g., Mihos \& Hernquist \cite{Mihos96}).

The
important aspect of star formation in a merger is related to
the processes taking place in the pre-merger phase.
As shown by Elmegreen et al. (\cite{Elmegreen93}), the interstellar gas
of pre-merger galaxy
is highly agitated. This results in a
significant
increase of the Jeans mass for forming gravitationally bound molecular
clouds
(up to 10$^{8}$--10$^{9}$ $M$\sunn).
The latter
could be progenitors of the HS 0837+4717 very massive starbursts.
As discussed, e.g., by Elmegreen
(\cite{Elmegreen99}), in regions with very high SF activity one should
expect a shift of the IMF to higher masses. Similarly, due to the lower
cooling efficiency at very low metallicity, the Jeans mass
is expected to be
higher, again resulting in an upward shift of the respective IMF.
This could result in a significant increase of the relative fraction of
massive stars, and
a larger effect of WR stars on
the local nitrogen enrichment.
Thus, the option of
merging of two slowly evolving XMD LSB galaxies could be a plausible
explanation for the XMD BCGs with such atypical properties.

\subsection{Summary and Conclusions}

We performed a complex study of one of the very metal-deficient BCG HS
0837+4717.
The long-slit high S/N ratio spectrophotometry was complemented by a
study of the ionized gas velocity distribution,
by \ion{H}{i} radio line 21-cm total flux measurements and by
broad-band
imaging photometry. The observed properties of this object are extreme
in several aspects.

This is the most luminous BCG with a metallicity less than or of the
order of $Z$\sunn/20. Its observed unusual colours are
caused mainly by the very strong [\ion{O}{iii}] emission lines shifted into
the
$V$-band due to its significant redshift. The factor of 6  nitrogen
overabundance in its supergiant \ion{H}{ii} region, in comparison to
other very metal-deficient BCGs, implies either a non-typical evolution
track for HS 0837+4717, or some very short transitional phase of star
formation, or both.
The ionized gas kinematics appears rather disturbed. There is an
indication of either counter-rotation, or of a supershell.
The BCG overall disturbed optical morphology with the "double-nucleus"
central structure, and two small appendages at its periphery, may
support the hypothesis of an advanced merger.
Recent images of this BCG from the SDSS DR1 database corroborate
the described appearance.

Among the known BCGs with the well measured N/O, there are two objects
with the larger N excesses, and three more BCGs with an excess of a
factor
of 3 relative to the norm for this type of objects. All of them are WR
galaxies; almost all belong to the group of luminous BCGs; and all
can be interpreted as various stages of mergers. This suggests that HS
0837+4717 could presumably be the most metal-poor representative of this
group.

From the results and discussion above we draw the following
conclusions:
\begin{enumerate}
\item
 The oxygen abundance in HS~0837+4717 is 12+$\log$(O/H)=7.64.
 The abundance ratios X/O for the elements Ne, Ar, S, Cl and Fe are
 consistent with the average values found for low-metallicity BCGs by
 Izotov \& Thuan (\cite{IT99}), implying the primary origin of these
 elements along with oxygen.
\item
 HS~0837+4717 is the most metal-deficient galaxy among the luminous BCGs
 with $M_{\rm B}\le$--18\fm0.
 This BCG with its O/H and $M_{\rm B}$ significantly deviates from the
 $Z$--$L_{\rm B}$  relationship derived for the large BCG sample.
\item
 We detected the broad emission features characteristic of WR
 stars, and estimated the number of WR stars as $\sim$1000.
 Their ratio to the number of O-stars is well compatible
 with that predicted by the current  models.
\item
 Broad low-contrast components of the H$\alpha$, H$\beta$, and
 [\ion{O}{iii}]~$\lambda\lambda$4959, 5007 emission lines were
 detected, with FWHMs of $\sim$1500 \kms.
 The flux ratio of the H$\alpha$ and H$\beta$ broad components
 implies very
 strong obscuration ($\sim$5 mag in $B$-band) in the emission region
 or along the line of sight. The extinction-corrected
 luminosity of the BCG should be higher than that observed, at least,
 by a factor of 2.5--5.
\item
 The position-velocity diagrams for the H$\alpha$ narrow component are
 not compatible with regular rotation. The major axis P--V diagram
 indicates
 either counter-rotation, or the presence of a giant supershell with a
 diameter of  $\sim$3.5  kpc and an expansion  velocity of about
 70~\kms.
 Half of the P--V diagram can be interpreted as rotation with an
 amplitude of $\sim$50--70~\kms\ at a radius of $\sim$4 kpc.
\item
 We find that this BCG is asymmetric both
 near the center (the inner structure consisting of two compact regions
 $\sim$2 kpc apart) and at the periphery. While the disturbed
 morphology of HS 0837+4717 could suggest a recent strong interaction,
 no candidate galaxies are found in its vicinity. Along with the very
 disturbed gas kinematics this could be evidence for a hypothesis of
 a recent merging of two gas-rich dwarfs.
\item
 The nitrogen-to-oxygen abundance ratio in HS 0837+4717 is six times
 higher than that of other XMD BCGs. The properties common to HS
 0837+4717 and five other non-XMD BCGs with the large nitrogen abundance
 excesses,  suggest that the large nitrogen overabundance could be
 connected with merger events and with a short phase of the related
 powerful starburst, when many WR stars contribute to the ISM enrichment.
\item
 HS 0837+4717 is located in a region with very low density of
 galaxies (void), the nearest of them with known redshift being situated at
 $\sim$6.4 Mpc~h$^{-1}$. The strong isolation of the BCG or its progenitor(s)
 might be the reason for its slow chemical evolution.
\end{enumerate}

\begin{acknowledgements}
We are grateful to the referee D.Kunth for useful suggestions,
 which allowed to improve the analysis and results presentation.
  The authors acknowledge the partial support from INTAS grants 96-0500
and
97-0033. SAO authors acknowledge the partial support from the Russian
Federal program "Astronomy". Y.I. thanks the support of National
Science
Foundation
grant AST 02-05785 and of the Swiss SCOPE 7UKPJ62178 grant. N.B.
acknowledges
support from the Israel Science Foundation and from Austrian Friends of
Tel Aviv University.
This research has made use of the NASA/IPAC Extragalactic
Database (NED), which is operated by the Jet Propulsion Laboratory,
California Institute of Technology, under contract with the National
Aeronautics and Space Administration.
The SDSS DR1 database is completed for public use thanks to the
funding by the Alfred P. Sloan Foundation, the SDSS member
institutions,
the National Aeronautics and Space Administration, the National Science
Foundation, the U.S. Department of Energy, the Japanese Monbukagakusho,
and
the Max Planck Society. The SDSS Web site is http://www.sdss.org/.
\end{acknowledgements}

\end{document}